# Noise-reduction techniques for $^1$H-FID-MRSI at 14.1T: Monte-Carlo validation & *in vivo* application

Brayan Alves[1,2*], Dunja Simicic[1,2,3*], Jessie Mosso[1,2,3], Thanh Phong Lê[3], Guillaume Briand[1,2], Wolfgang Bogner[4], Bernard Lanz[1,2], Bernhard Strasser[4], Antoine Klauser,[5#], Cristina Cudalbu[1,2#]

[1] CIBM Center for Biomedical Imaging, Switzerland

[2] Animal Imaging and Technology, École polytechnique fédérale de Lausanne (EPFL), Lausanne, Switzerland

[3] Laboratory of Functional and Metabolic Imaging, École polytechnique fédérale de Lausanne (EPFL), Lausanne, Switzerland

[4.] High-field MR Center, Department of Biomedical Imaging and Image-guided Therapy, Medical University Vienna, Vienna, Austria

[5] Advanced Clinical Imaging Technology, Siemens Healthcare AG, Lausanne, Switzerland.

*Joint first authors

#Joint last authors

**Abbreviations:**

MRSI - Magnetic Resonance Spectroscopic Imaging, FID - Free Induction Decay, AD - Acquisition Delay, TR - Repetition Time, MP-PCA - Marchenko-Pastur Principal Component Analysis, LR-TGV - Low-Rank Total Generalized Variation, SVD - Singular Value Decomposition, SNR - Signal to Noise Ratio, SD - Standard Deviation, FWHM - Full Width at Half Maximum, GT - Ground Truth, MC - Monte-Carlo, NAA - N-Acetyl Aspartate, NAAG - N-acetylaspartylglutamate, Ins - myo-Inositol, Gln - Glutamine, Glu - Glutamate, GABA - Gamma-Aminobutyric Acid, Asp - Aspartic Acid, Ala - Alanine, Asc - Ascorbic Acid, Cr - Creatine, PCr - Phosphocreatine, tCr - total Creatine, Tau - Taurine, PCho - Phosphocholine, GPC - Glycerophosphocholine, PE - Phosphoethanolamine, Lac - Lactate, Glc - Glucose, GSH - Glutathione

# Abstract

Proton magnetic resonance spectroscopic imaging ($^1H$-MRSI) is a powerful tool that enables the multidimensional non-invasive mapping of the neurochemical profile at high-resolution over the entire brain. The constant demand for higher spatial resolution in $^1H$-MRSI led to increased interest in post-processing-based denoising methods aimed at reducing noise variance.

The aim of the present study was to implement two noise-reduction techniques, the Marchenko-Pastur principal component analysis (MP-PCA) based denoising and the low-rank total generalized variation (LR-TGV) reconstruction, and to test their potential and impact on preclinical 14.1T fast *in vivo* $^1H$-FID-MRSI datasets. Since there is no known ground truth for *in vivo* metabolite maps, additional evaluations of the performance of both noise-reduction strategies were conducted using Monte-Carlo simulations.

Results showed that both denoising techniques increased the apparent signal-to-noise ratio SNR while preserving noise properties in each spectrum for both *in vivo* and Monte-Carlo datasets. Relative metabolite concentrations were not significantly altered by either methods and brain regional differences were preserved in both synthetic and *in vivo* datasets. Increased precision of metabolite estimates was observed for the two methods, with inconsistencies noted on lower concentrated metabolites.

Our study provided a framework on how to evaluate the performance of MP-PCA and LR-TGV methods for preclinical $^1H$-FID MRSI data at 14.1T. While gains in apparent SNR and precision were observed, concentration estimations ought to be treated with care especially for low-concentrated metabolites.

# 1. Introduction

Proton magnetic resonance spectroscopic imaging ($^1H$-MRSI) is a powerful tool that enables the *in vivo* acquisition of MR spectra from multiple spatial positions simultaneously. This technique enables non-invasive mapping of neurochemical profile at high-resolution over the entire brain[1,2]. Free induction decay $^1H$-MRSI ($^1H$-FID-MRSI) sequences have seen an increased usage in clinical settings at ultra-high field[3–6], leading to successful investigations of various pathologies such as glioma or multiple sclerosis[7–10]. To the authors' knowledge, $^1H$-FID-MRSI is not routinely used in the preclinical context due to several challenges, in particular the lack of sensitivity due to low metabolite concentrations and small nominal voxel volumes in rodents resulting in long acquisition times, as well as advanced pulse sequences and processing methods that need to be implemented in-house[5,6]. Furthermore, signal-to-noise ratio (SNR) issues can reduce quantification precision, which is paramount to measure brain regional differences. The constant demand for higher spatial resolution is leading to increased interest in post-processing-based denoising methods aimed at reducing noise variance in $^1H$-MRSI[11–17].

Reconstruction techniques that leverage various low-rank (LR) assumptions have been widely applied and implemented within the clinical settings[14,16]. These methods rely on linear predictability and/or partial separability of spatial-temporal modes to denoise MRSI data[12–14]. Low-rank reconstruction methods can be combined with constraints on the signal spatial distribution with specific regularization to further improve SNR in MRSI reconstruction[16,18–20]. For example, in Klauser et al.[14] partial separability is combined with a total generalized variation (TGV) spatial regularization in a reconstruction model (LR-TGV) to preserve edges while avoiding stair-casing artifacts[21].

In parallel, the Marchenko-Pastur principal component analysis (MP-PCA) based denoising has been implemented for diffusion MRI[17], functional MRI[22,23], and diffusion-weighted MRS[24,25] (DW-MRS) and was recently applied to preclinical $^1H$-MRSI at 9.4T[15]. MP-PCA technique exploits the fact that noise eigenvalues follow the universal Marchenko-Pastur distribution, a result of the random matrix theory. This method provides a data-driven approach to distinguish noise from signal components. Moreover, the performance of MP-PCA denoising can be improved with matrix size and spectral redundancy throughout the image, which is particularly relevant for high-resolution MRSI[15].

The aim of the present study was to implement these two noise-reduction techniques, MP-PCA based denoising and LR-TGV reconstruction, and to test their potential and impact on preclinical 14.1T fast *in vivo* $^1H$-FID-MRSI datasets. Since there is no known ground truth for *in vivo* metabolite maps, additional performance evaluations of both noise-reduction strategies were conducted using Monte-Carlo simulations.

## 2. Methods

For common understanding, the following terminology is adopted throughout this manuscript. *Model FID* is used in the Monte-Carlo study to reference a noiseless free induction decay (FID) manually generated with a linear combination of individual metabolites of interest simulated via NMR ScopeB[26] and *in vivo* acquired macromolecules (Figure 1A). *Signal-to-noise ratio* (*SNR*) corresponds to the spectral SNR calculated from the height of the NAA singlet at 2.01 ppm in the magnitude spectrum, divided by one standard deviation (SD) of the noise of the real spectral component measured in a noise-only region (from 0.54 to (-0.98) ppm)[25]. The term *apparent SNR* will be used to refer to the SNR after denoising. *Relative concentration* corresponds to the voxel-wise ratio of the concentration output by the internal reference total creatine (tCr) (fixed at 8 mmol/kg$_{ww}$), calculated by LCModel[27]. *Noise distribution* refers to the intensity distribution of the real spectral component in a noise-only region. *Spectral residuals* correspond to the difference of the real spectral component between two MRSI datasets, in a metabolite-containing region, where spatial dimensions are concatenated such that one dimension corresponds to the number of spectral points and the other to the voxel index. The data framework used for this study as well as the description of both MP-PCA and LR-TGV can be found in Supplementary Materials sections 1-3.

### 2.1. Sequence

The 2D fast $^1$H-FID-MRSI protocol used for *in vivo* acquisitions was implemented on a 14.1T Bruker scanner as described by Simicic et al.[28]. Data were acquired using a homemade quadrature surface coil (1.8×1.6 cm inner loop radius). The sequence used a Shinnar–Le Roux excitation pulse with an 8.4 kHz bandwidth adjusted to the 52° Ernst angle with an acquisition delay AD=1.3 ms and a repetition time TR=813 ms. A Cartesian *k*-space sampling was used with a 7.143 kHz acquisition bandwidth and 1024 spectral points. A 2 mm-thick coronal slice with a 24×24 mm$^2$ field-of-view, was centered in the brain to cover the hippocampus, a part of striatum and cortex, based on anatomical images acquired with T$_2$-weighted Turbo-RARE on the axial and coronal directions (256×256 resolution, 20 slices, RARE$_{factor}$=6). A 31×31 matrix was used, leading to a nominal 0.77×0.77×2.00 mm$^3$ voxel size. MRSI acquisitions were performed with 1 average (standard mode, no post-processing filters applied). The sequence used a VAPOR water suppression scheme for metabolite acquisitions. Shimming was performed using Bruker MAPSHIM first in the full brain and then in a 10×10×2 mm$^3$ volume of interest centered in the MRSI slice. Six saturation slabs were placed around the volume of interest to minimize lipid contamination. Overall MRSI acquisition time was 26 minutes, for both water and metabolites acquisitions combined (i.e. 13 min each). More information on the application of this protocol can be found in[28] and [LIVE Demos – MRS4BRAIN - EPFL](). Parameters used for this study can be found in Supplementary Materials Section 4, Table S1.

## 2.2. *In vivo* Experiments

MRSI data were acquired on five Wistar rats (250-300 g, males, Charles River laboratories, L'Arbresle, France) under isoflurane anesthesia (1.5-2.0% in 60% oxygen/40% air, respiration rate maintained at 60–70 breaths/min). For the MR measurements, animals were placed in an in-house built holder, and their head was fixed in a stereotaxic system using a bite bar and a pair of ear bars. Respiration rate and body temperature were monitored using a small-animal monitoring system (SA Instruments, New York, NY, USA). Body temperature was measured with a rectal thermosensor and maintained at 37.7±0.2°C by a warm water hose circulation. All animal experiments were conducted according to federal and local ethical guidelines, and the protocols were approved by the local Committee on Animal Experimentation for the Canton de Vaud, Switzerland (VD 3022.1).

## 2.3. Monte-Carlo Simulations

In order to assess the performance of both noise-reduction strategies with a known ground-truth, synthetic MRSI slices were generated using a homemade MATLAB script (MathWorks, Natick, MA, USA). The anatomical model chosen for the slice was inspired by *in vivo* acquisitions using an ITK-Snap brain mask created from a coronal image acquired during the *in vivo* study. Supplementary Material Section 5 describes how the number of Monte Carlo simulations was selected.

A first proof of concept simulation performed with 2-compartments is reported in Supplementary Materials Section 6, Figures S1 and S2. It has been previously shown that the presence of B0 shifts affects the number of components retained by the MP fit for diffusion weighted MRS data[25]. As such, for this proof of concept, the impact of applying a Gaussian distributed $B_0$-field map or a realistic *in vivo* $B_0$-field map (see below for details) on concentration estimates and their brain regional differences was assessed.

Then, a 3-compartment geometry was used to schematically depict the hippocampus, the striatum and the cortex (Figure 1B) containing only the realistic $B_0$-field map. Each compartment contains a model FID generated from the basis set components (19 metabolites and macromolecules, see Figure 1A) multiplied by a factor $\alpha(r, t)$ representing the local *in vivo* reported concentration in each of the three selected brain regions[29] (values reported in the Supplementary Materials Section 7, Table S2), for a total of 317 model FIDs (distribution reported in Figure 1B). By that, the rank of the ground truth data without perturbations like noise or $B_0$-inhomogeneity is three. A Gaussian distributed noise amplitude was added to each model FID (NAA linewidth of 17.4, 17.8 and 10.6 Hz in compartment 1, 2 and 3, respectively) with a fixed SD for the whole slice. The noise amplitude $\sigma$ was computed from the ratio between the highest amplitude point of the real FID component and the input SNR level chosen by the user. In the scope of this work, multiple input SNR levels were simulated (input SNR$\in\{3,5,7,10,12,15,20\}$ corresponding a posteriori to LCModel SNR$\in\{3,4,6,9,10,13,17\}$) using 30 Monte-Carlo realizations per SNR with random noise generations each time.

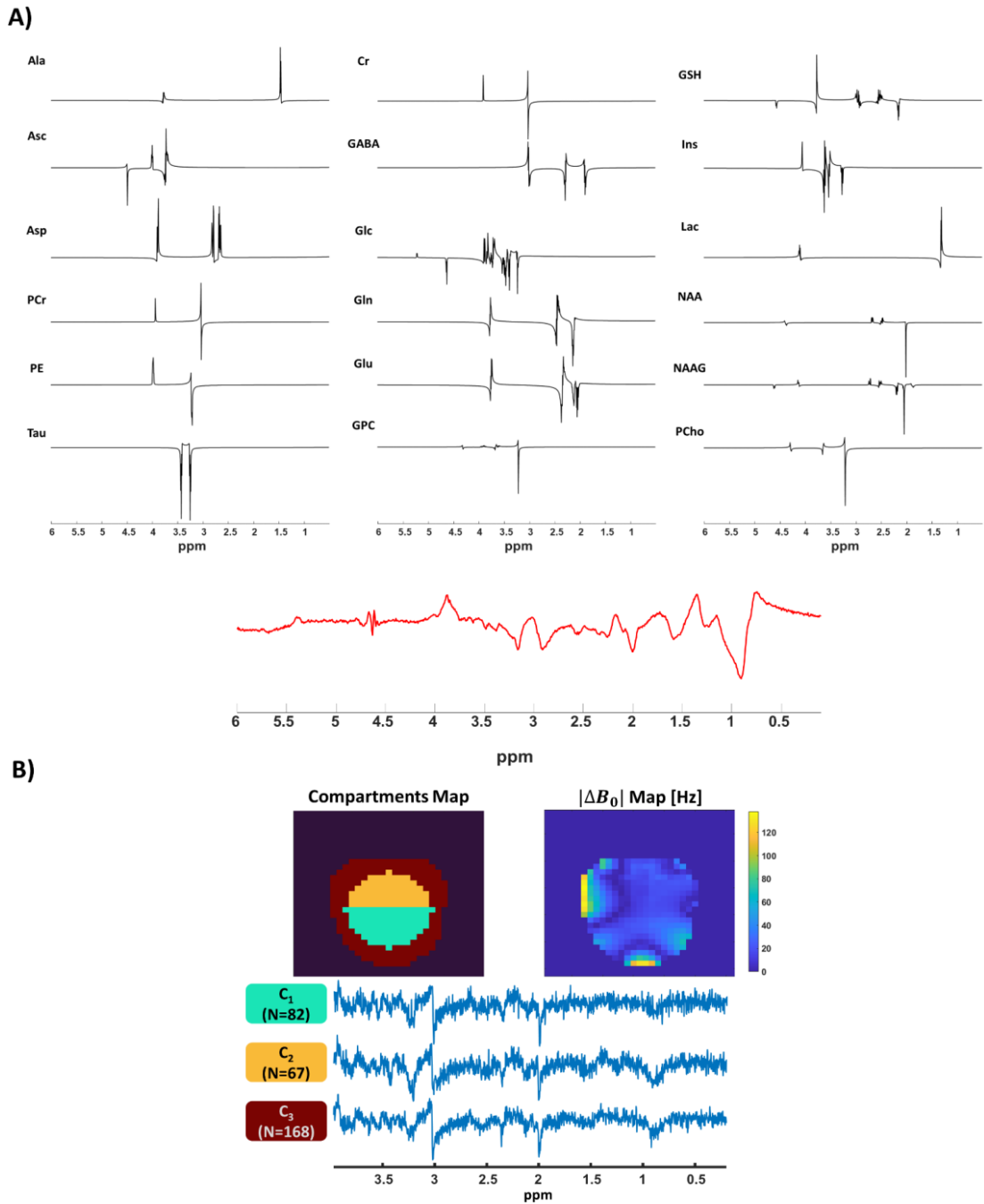

*Figure 1 A) Basis set elements (metabolites) simulated in NMR-Scope B (black) and macromolecules (MM, red) acquired in vivo. B) Monte-Carlo maps of the distribution of the simulated spectra and of the applied in vivo $\Delta B_0$. The 3 different simulated spectra distributed in the synthetic MRSI slice are displayed with an SNR of 7. Representative spectra for all input SNR can be observed in Supplementary Material Section 7, Figure S3.*

The lower SNR levels correspond to SNR values obtained *in vivo*, while the higher levels match the apparent SNR values measured after application of either denoising methods on *in vivo* datasets. The effect of realistic static field inhomogeneities were simulated on each FID with respect to a $B_0$-field map acquired *in vivo* (Figure 1B), by multiplying the voxel corresponding model FID with a shift term $e^{-2\pi it\Delta B_0(r)}$.

The synthetic MRSI slice was processed in a customized version of the processing pipeline described below (section 2.4). Each dataset was triplicated: one dataset was denoised using LR-TGV, one dataset with MP-PCA (for clarity, they will be named LR-TGV data and MP-PCA data) and one dataset was kept as is (called RAW data). For the Monte-Carlo study, only denoising and quantification were performed (steps 4 to 7 in processing pipeline). In order to eliminate any potential confounding factors due to systematic under/overestimation of concentrations with LCModel for raw data, the ground truth (GT) used for the Monte-Carlo (MC) study was generated using a synthetic set with a sufficiently high SNR to be considered as noiseless data (fixed SNR at 300). The evaluation of the concentration in each compartment was averaged over the number of compartment voxels and the number of MC realizations.

## 2.4. Data Processing and Denoising

A homemade MATLAB processing pipeline built on the work of Klauser et al.[14] was prepared for *¹H-FID-MRSI* datasets containing the following steps (sketch in Figure 2):

1. Removal of the residual water signal using a Hankel-Singular Value Decomposition (HSVD) water removal[14,30,31].
2. SVD-based lipid removal was applied on the raw dataset following the methodology proposed by Klauser et al.[14]. This fast method assumes that lipids and metabolites are orthogonal in the temporal domain as demonstrated by Bilgic et al.[32,33]. To apply this method, a brain/scalp segmentation based on the water power mask was applied using the power of the acquired water dataset. The water power signal was computed by summing the squared magnitude of the frequency-domain signal. Considering that the scalp region is covered by saturation bands, a segmentation criterion for the water power mask was defined: voxels having a power greater than half the average power in the slice were considered as being within the brain region (Figure 2C). Using signals originating from scalp voxels, it was then possible to generate an orthogonal basis for the approximated lipid subspace using SVD. The rank of the basis was determined by a threshold on the mean power ratio between the brain and the scalp region after application of the lipid suppression: $E_{Brain}/E_{scalp} \geq \alpha$ (more information in 15), with α being manually set by the user in the pipeline, 0.5 in our study.
3. Triplication of each dataset (Figure 2A).

4. The LR-TGV denoising was applied to the full dataset (31×31 matrix). The corresponding water acquisition was used as input for $B_0$-field map calculation. As the number of components $N_{rank}$ for LR-TGV is determined empirically, in the present study, $N_{rank}$ was matched with the MP-PCA estimation in the same datasets. Using the same rank for both methods ensures a fair comparison of the results. $N_{rank}$ for the Monte-Carlo sets was based on MP-PCA denoising: each set was firstly denoised by the MP-PCA method and the number of components found by MP-PCA was then used as rank for the LR-TGV denoising (for the corresponding 6 input SNR, averaged over 30 repetitions: $N_{rank} \in \{9,11,12,13,14,15,17\}$).

5. For the MP-PCA, only the voxels within the water power mask were processed. MP-PCA was applied with centering and no moving/sliding window[25]. No corrections for $B_0$ shifts were performed. The complex-valued FIDs were split into real and imaginary parts, where the first dimension contained the time-domain sampling (1024 points) and the second dimension the number of spectra selected. A truncated PCA was then performed.

6. The resulting spectra: RAW, LR-TGV denoised and MP-PCA denoised were quantified with LCModel (Version 6.3-1N) using a metabolite basis-set simulated in NMR ScopeB using published values of J-coupling constants and chemical shifts and the pulse-acquire sequence with the same parameters as for the *in vivo* $^1H$-FID-MRSI acquisitions, combined with an *in vivo* measured macromolecules spectrum[28]. Zero-order phase correction was performed by LCModel. The elements of the metabolite basis-set can be seen in Figure 1A.

7. The water power mask segmentation was used to exclude the voxels located outside the brain, while on the remaining voxels two semi-automatic quality control criteria were applied to keep only the reliably quantified spectra. As a first step for quality control assessment, the pipeline uses the LCModel values of SNR and FWHM and averages both over the number of voxels: $\underline{SNR}$ and $\underline{FWHM}$ Fixed quality thresholds were chosen to be set above 75% of $\underline{SNR}$ and below 125% of $\underline{FWHM}$. The average values $\underline{SNR}$ and $\underline{FWHM}$ used for these thresholds were computed from the RAW dataset and then applied on the three sets (Figure 2C). The metabolite maps were overlaid onto the corresponding anatomical MRI image.

8. *In vivo*, a segmentation of two different brain regions was performed using the *MRS4Brain toolbox*[34]: a region with a mix of striatum and cortex and the hippocampus to evaluate whether the brain regional differences can be retrieved by both denoising techniques. An average concentration was computed by averaging over the voxels contained in the specific region. An additional acceptance criterion with respect to the Cramer-Rao lower bound (CRLB) of the voxel on the RAW dataset was added for the *in vivo* datasets: for each metabolite, the CRLB had to be lower or equal than 40%. Approximately 30 and 34 voxels were selected after application of all acceptance criteria for the hippocampus and striatum+cortex region, respectively. As the CRLB may not be a valid metric to distinguish reliable metabolite estimation for denoised sets[13,25], the set of voxels selected in the RAW dataset was used for both MP-PCA and LR-TGV datasets.

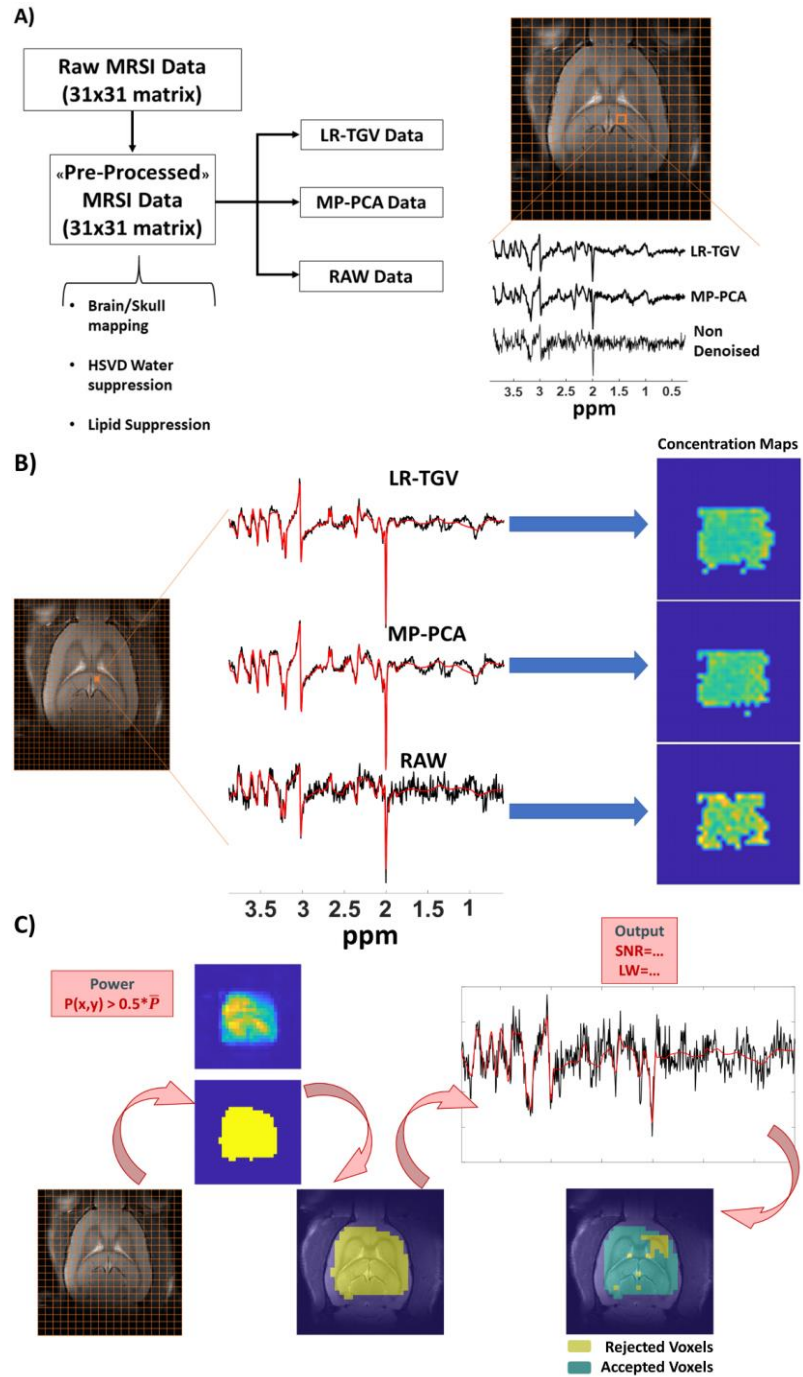

*Figure 2 A) Sketch of the processing pipeline used for this study. B) Output in vivo spectra after LCModel quantification. The concentration results are then transferred in 31×31 matrices to represent the MRSI slice. C) Quality control process. The first step uses the water power mask calculated during pre-processing. The SNR and FWHM (LW-linewidth) criteria estimated by LCModel are then used to generate a "quality control" mask.*

## 2.5. Statistical Analysis

Absolute and relative concentrations are presented as mean±SD across regions and realizations for the MC study and across regions and datasets for the *in vivo* study. A two-way analysis of variance (two-way ANOVA) was performed with respect to each metabolite using GraphPad Prism version 5.04 for Windows (GraphPad Software, San Diego, California USA) followed by Bonferroni's multi-comparison post-hoc tests. For statistical analysis, two categorical factors were defined for the methodology (RAW, LR-TGV, MP-PCA) and for the brain regions (Cortex+Striatum and Hippocampus). All tests were two-tailed. Significance in Bonferroni post-hoc tests were attributed as follows: $*p < .05$; $**p < .01$; $***p < .001$; $****p < .0001$.

# 3. Results

The impact and performance of the two noise-reduction strategies were evaluated by assessing the noise distribution, spectral residuals between RAW vs. MP-PCA and RAW vs. LR-TGV, LCModel apparent SNR, map coverage, precision (SD), accuracy and variations of metabolite concentration estimates. For the MC study, metabolite concentration estimates in each compartment were assessed for various input SNR and compared against the ground truth at SNR=300 for accuracy evaluation.

The effect of both denoising techniques on the metabolite map coverage and preservation of brain regional difference in metabolite concentration estimates was evaluated by computing: 1) the fraction of voxels accepted by the semi-automatic quality control; and 2) the average relative concentration of Ins, Gln, Glu, GABA and NAA in the hippocampus and cortex+striatum region, to assess well known brain regional difference after denoising.

## 3.1. Noise properties

Figure 3I.A and 3II.B show that the noise from the RAW dataset follows a Gaussian distribution. This distribution was preserved regardless of the applied noise-reduction method as shown by both QQ-plots. It is worth noting that both denoising techniques led to a uniform noise-reduction across the MRSI slice for *in vivo* datasets, while in the Monte-Carlo sets, voxels around the border of the map had a lower reduction of the noise level (Supplementary Material Section 7, Figure S3). Figure 3I.B and 3II.B illustrate the difference between the raw data and denoised datasets for both techniques, with a certain pattern observed for the LR-TGV residuals around 2 ppm and 3 ppm. This pattern does not appear for the MP-PCA residuals, highlighting a uniform noise level across spectral points and a uniform noise level across MRSI FIDs inside the matrix. The stack plots in Figure 3I.C and 3II.C further emphasize this specific pattern around 2 and 3 ppm.

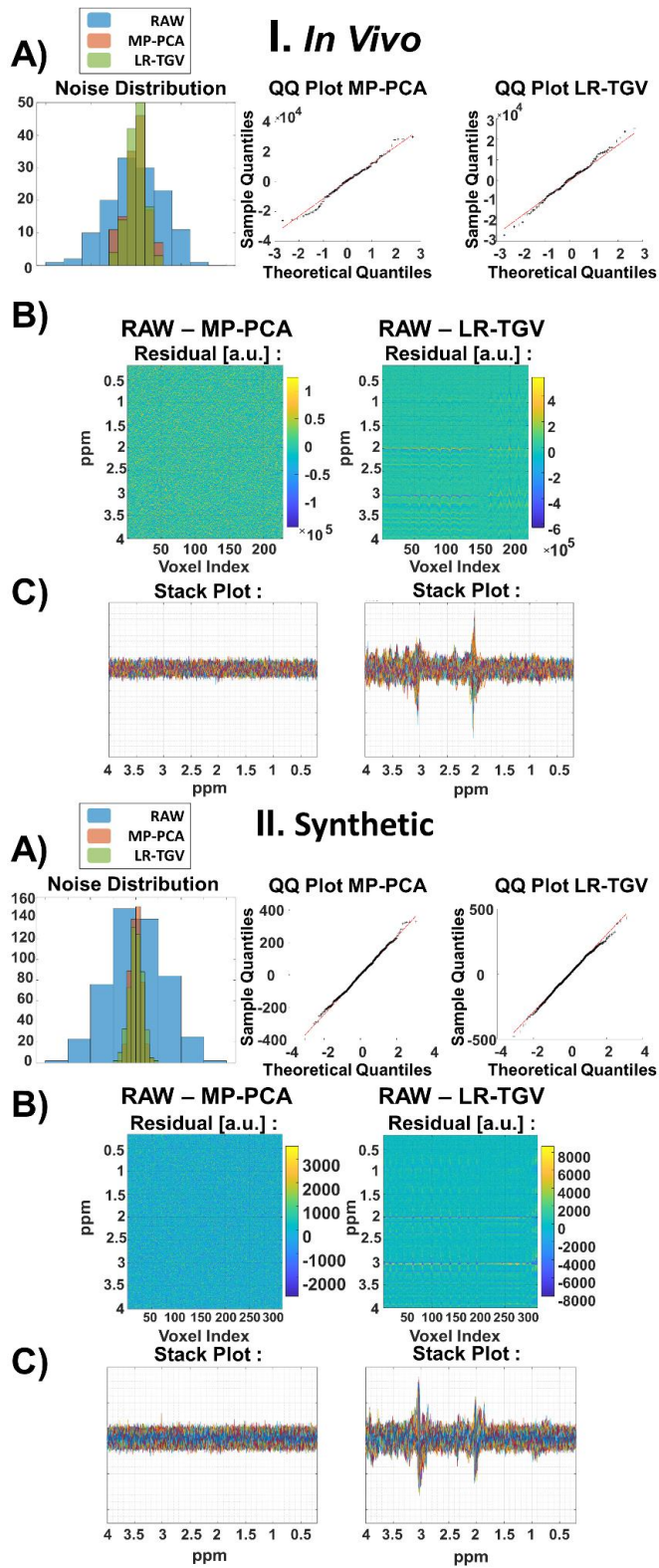

*Figure 3 Noise properties of the in vivo (I.) and synthetic (II.) sets. A) Noise distribution for the three sets of data as well as QQ plots of the MP-PCA and LR-TGV sets; B) Residuals between raw and denoised data. The difference was computed for the metabolite part of the spectral range (from 4 to 0 ppm). Rows: spectral points. Column: voxel index (starting from the top left to the bottom right corner, following the water power mask line by line); C) Stack plots of the residual spectra for both methods. Both denoising methods lead to an increased apparent SNR and preserve the Gaussian noise distribution. MP-PCA*

*denoising revealed no structure in spectral residuals while for LR-TGV some spectral residuals can be observed (395% and 381% increase compared to the residual amplitude for in vivo and MC respectively).*

### 3.2. Monte-Carlo simulations

The LCModel apparent SNR after each denoising method was consistently increased for each input SNR values (apparent $SNR_{MP-PCA} \in \{13,19,26,32,39,41,51\}$, apparent $SNR_{LR-TGV} \in \{12,24,34,34,55,60,60\}$ for corresponding input $SNR \in \{3,5,7,10,12,15,20\}$). No alteration of the linewidth was observed with either noise-reduction techniques. Representative spectra for raw, MP-PCA and LR-TGV denoised datasets can be seen on Supplementary Material Section 7, Figures S3-5.

The 2-compartments model was combined with a Gaussian distributed $B_0$-field map at two different SNR values (7 and 15) and for a majority of metabolites, a loss of regional difference was observed. The regional difference was recovered when the realistic $B_0$-field map seen in Figure 1B was applied in the same 2-compartment model (Supplementary Material Section 6, Figure S2).

The results of 3-compartments simulations are displayed in Figure 4A-B. The concentration maps for NAA, GABA and Gln were chosen as representative examples due to their different concentration levels (NAA high concentration, Gln and GABA low concentration and J-coupled) and their spatial distribution. Less metabolite variation was observed within a compartment for the two denoised sets compared to the RAW set, with reduced average concentration SD at low SNR per compartment (bar plot on Figure 4C). Furthermore, both denoising techniques led to concentration estimates close to the GT, thus preserving the regional differences.

For Gln, an average increase in precision of 68% and 64% over the three compartments was observed for MP-PCA at input SNR=5 and 12, respectively (Figure 4C). The Gln precision for LR-TGV increased by 76% and 69% on average for an input SNR=5 and 12, respectively. Regional differences were consistently retrieved for the RAW set at all input SNRs. The application of the two noise-reduction techniques did not alter these differences.

For NAA, where the compartments $C_1$ and $C_3$ (79% of the brain mask) have very similar concentration compared to $C_2$ (21% of the brain mask), the differences were also preserved. The precision averaged over the three compartments for MP-PCA was increased by 45% and 20% for both input SNR=5 and 12, while it was increased by 60% at input SNR=5 and 39% at input SNR=12 for LR-TGV.

For GABA, a low concentration, overlapping and J-coupled metabolite which was not well quantified for input SNR=5 in the RAW data, the differences between compartments were recovered by the two denoising techniques. However, the individual concentrations were not fully recovered, although closer to the GT after denoising compared to RAW data. Precision averaged over the three compartments was increased by 46% with MP-PCA and by 62% with LR-TGV at input SNR=5. For input SNR=12, a higher increase of the precision was noted with a 63% increase for the MP-PCA and a 72% increase for LR-TGV.

The same procedure for the estimation of metabolite concentrations was done for all input SNR to observe the behavior of the noise-reduction techniques at different input SNR (Figure 5). Both MP-PCA and LR-TGV techniques led to concentrations close to the GT, and consequently preserved the regional differences for all input SNR between 5 and 20. Furthermore, in the range of low SNR values, the two noise-reduction sets were closer to the GT than the RAW set. This result was consistently observed for all metabolites of interest, as can be seen in Table 1 and Supplementary Materials Figures S7-8. The former shows the accuracy of the relative concentration measured in RAW, MP-PCA and LR-TGV, averaged over the input SNR∈{3,5,7,10} representing the SNR values that are found on *in vivo* MRSI slices. It is worth noting that the SD was bigger in compartment 3 *vs* compartments 1 and 2 for the LR-TGV method.

The absolute concentration and relative regional difference compared to GT for different metabolites can be seen in Tables 1 and 2 respectively. All metabolites which were correctly quantified from RAW sets, were also correctly estimated after denoising. Moreover, lower concentration metabolites such as Asp or GABA were also evaluated correctly with denoising, while for the RAW data, the concentrations were not accurately estimated.

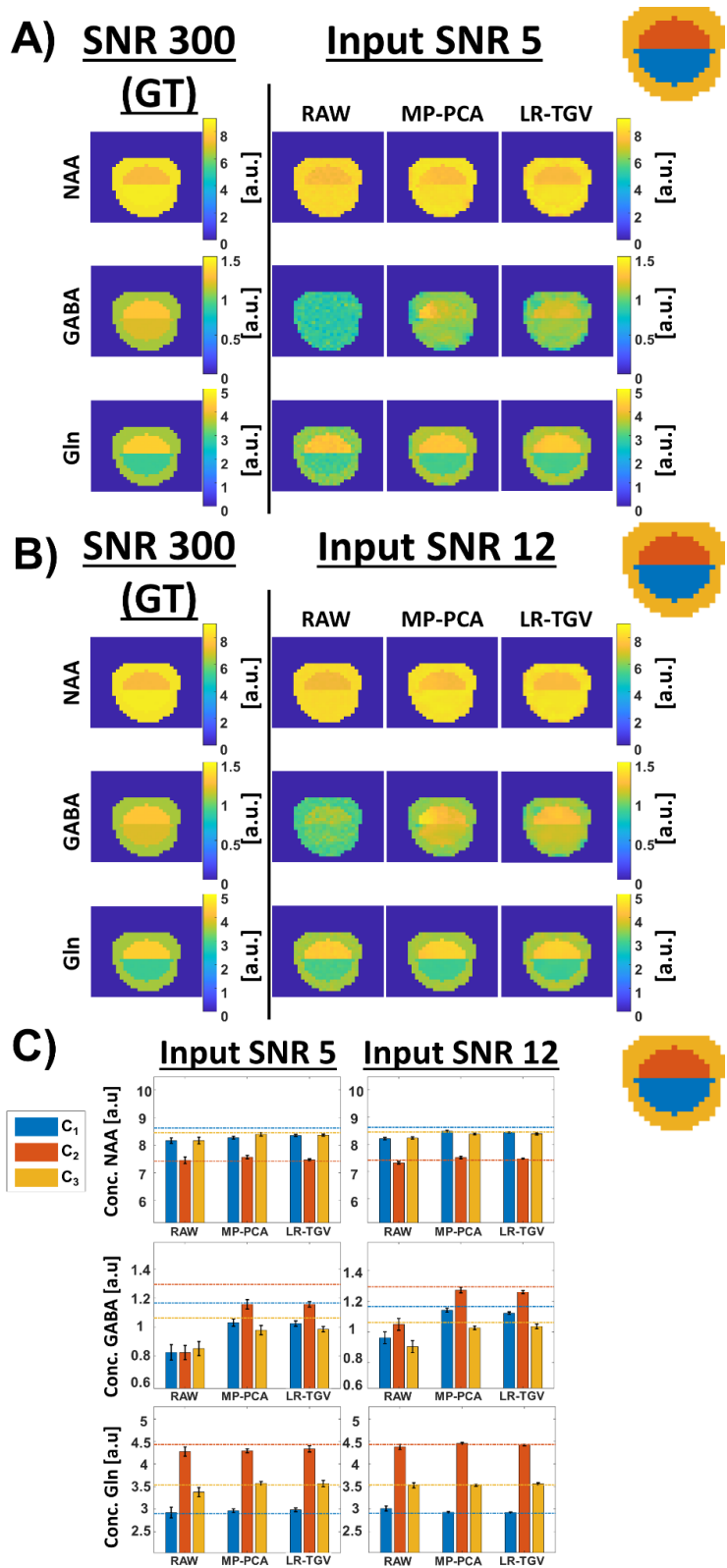

Figure 4 Synthetic concentration maps (NAA, GABA, Gln, in a.u. based on concentrations reported in Supplementary Table S2) A) at input SNR 5 and B) at input SNR 12 (representing LCModel apparent SNR 4 and 10, respectively). Ground truth (GT) map at input SNR=300 is displayed for comparison. C) Bar plot of the average concentration on each compartment, at input SNR 5 and 12. GT at SNR=300 is represented in dotted lines on the bar plot.

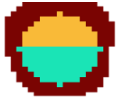
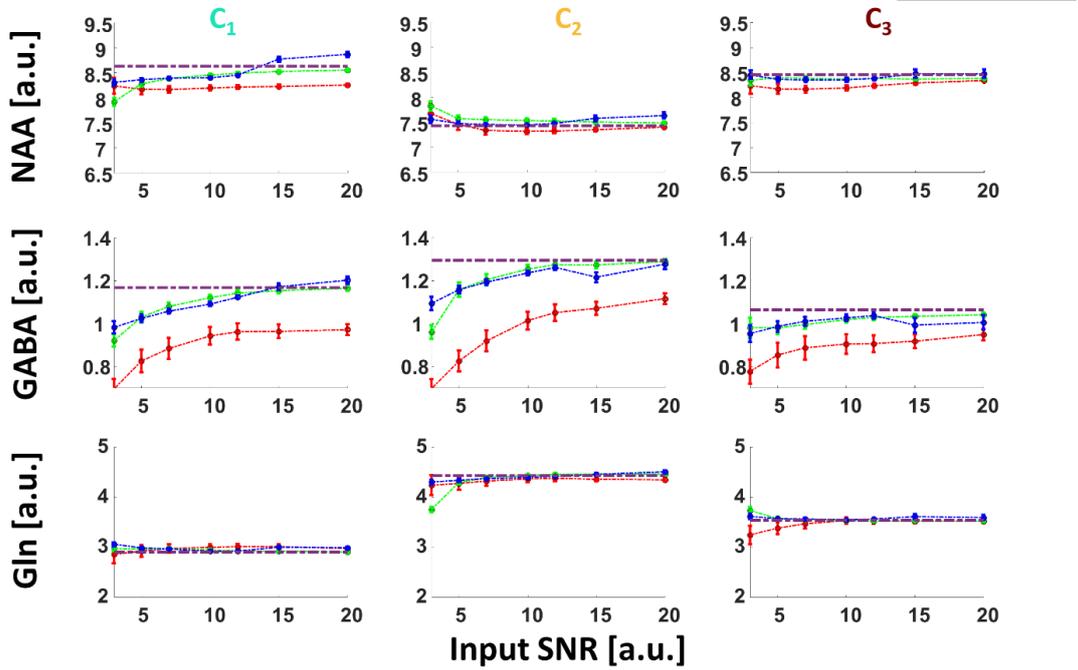
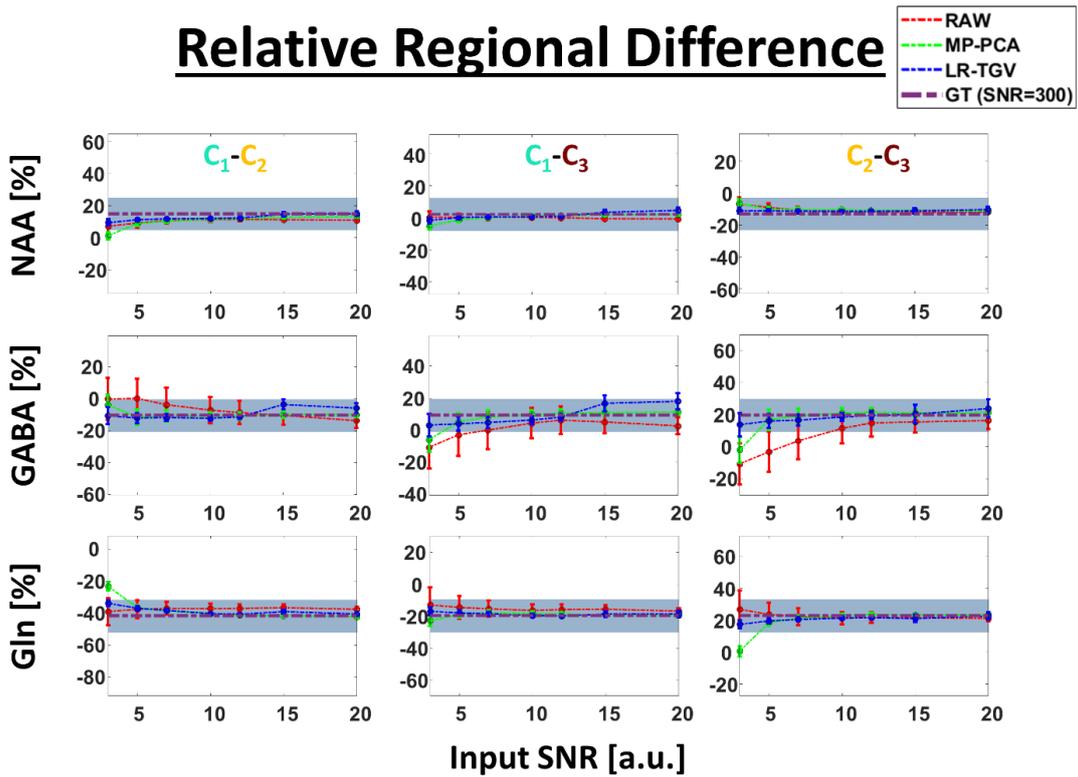

*Figure 5 Plots of the absolute concentration and relative regional difference as a function of the input SNR. The ground truth value of concentration and difference is displayed in a dotted purple horizontal line. The blue region represents the 10% difference range around the ground truth.*

| Relative Difference [%] | C1 | | | C2 | | | C3 | | |
|---|---|---|---|---|---|---|---|---|---|
| | RAW | MP-PCA | LR-TGV | RAW | MP-PCA | LR-TGV | RAW | MP-PCA | LR-TGV |
| Ala | 28.2 | 13.1 | 13.8 | 17.5 | 6.3 | 9.4 | 8.5 | 7.4 | 7.5 |
| Asp | 11.5 | 3.2 | 5.0 | 19.4 | 11.6 | 6.5 | 2.4 | 2.5 | 1.8 |
| GABA | 26.1 | 9.2 | 9.5 | 30.5 | 9.7 | 8.3 | 18.7 | 5.9 | 5.7 |
| Gln | 1.7 | 1.9 | 2.3 | 2.6 | 3.8 | 1.5 | 2.9 | 1.5 | 1.1 |
| Glu | 3.8 | 3.9 | 2.5 | 2.3 | 1.4 | 1.5 | 3.9 | 2.5 | 2.5 |
| GSH | 10.3 | 6.9 | 3.3 | 12.3 | 8.5 | 1.4 | 32.4 | 5.9 | 7.5 |
| Ins | 4.2 | 4.3 | 3.2 | 2.1 | 1.1 | 0.1 | 5.3 | 4.2 | 3.3 |
| NAA | 5.0 | 3.7 | 2.8 | 0.0 | 2.4 | 0.8 | 3.0 | 1.0 | 0.9 |
| Tau | 5.5 | 5.8 | 1.6 | 0.8 | 7.5 | 1.3 | 5.8 | 0.6 | 2.1 |
| NAAG | 25.7 | 13.9 | 16.2 | 5.1 | 5.4 | 6.7 | 7.6 | 4.8 | 3.7 |
| GPC+PCho | 12.6 | 3.9 | 7.8 | 13.6 | 2.6 | 4.1 | 13.4 | 7.9 | 7.9 |
| NAA+NAAG | 2.3 | 2.2 | 1.2 | 0.0 | 2.7 | 1.3 | 2.0 | 0.3 | 0.4 |

*Table 1 Absolute relative difference of the average concentration, with respect to the ground truth SNR=300, for all individual compartments. Values of concentrations are averaged over SNR ∈{3,5,7,10}. Differences larger than 10% are highlighted in red*

| Relative Difference [%] | C1-C2 | | | C1-C3 | | | C2-C3 | | |
|---|---|---|---|---|---|---|---|---|---|
| | RAW | MP-PCA | LR-TGV | RAW | MP-PCA | LR-TGV | RAW | MP-PCA | LR-TGV |
| Ala | 9.6 | 7.7 | 5.0 | 15.6 | 6.2 | 7.2 | 6.2 | 3.3 | 3.0 |
| Asp | 9.2 | 15.7 | 11.5 | 14.1 | 5.5 | 6.4 | 19.4 | 10.2 | 4.9 |
| GABA | 6.5 | 1.8 | 1.3 | 10.1 | 4.4 | 4.1 | 16.6 | 5.6 | 2.8 |
| Gln | 4.2 | 5.7 | 3.6 | 4.6 | 1.6 | 1.2 | 1.6 | 5.7 | 2.6 |
| Glu | 1.5 | 2.6 | 1.1 | 0.4 | 1.5 | 0.4 | 1.5 | 1.2 | 1.1 |
| GSH | 3.5 | 5.6 | 1.9 | 17.4 | 12.4 | 10.4 | 15.4 | 14.2 | 8.1 |
| Ins | 2.4 | 5.3 | 3.2 | 1.2 | 1.2 | 0.4 | 3.2 | 5.2 | 3.4 |
| NAA | 5.1 | 6.2 | 3.6 | 2.0 | 2.9 | 2.0 | 3.0 | 3.3 | 1.6 |
| Tau | 4.7 | 5.9 | 0.8 | 0.4 | 6.4 | 3.7 | 4.7 | 10.9 | 3.2 |
| NAAG | 17.9 | 10.9 | 8.3 | 15.6 | 8.2 | 11.2 | 5.9 | 7.2 | 3.0 |
| GPC+PCho | 1.9 | 6.5 | 3.3 | 2.1 | 3.9 | 0.7 | 4.0 | 10.1 | 3.4 |
| NAA+NAAG | 2.4 | 4.8 | 2.5 | 0.3 | 1.9 | 0.8 | 2.0 | 2.9 | 1.7 |

*Table 2 Relative regional difference between compartments, with respect to the ground truth SNR=300, for all regional differences available. Values of concentrations are averaged over SNR ∈{3,5,7,10}. Difference larger than 10% are highlighted in red*

## 3.3. *In vivo* experiments

Figure 6 shows the impact of both denoising strategies on brain coverage for *in vivo* rat brain data using NAA, Glu, Gln and GABA as examples. The coverage of the concentration maps increased with both denoising techniques; more voxels located around the edges passed the SNR criterion. It is worth noting that the usage of surface coils leads to a reduction of SNR at the edges of the MRSI slice. Therefore, this increase in coverage can be explained by the higher apparent SNR due to the noise-reduction. This is highlighted through the table in Figure 6, containing the averaged apparent SNR and FWHM over the slice and the percentage of voxels accepted after applying the quality control criteria. The average apparent SNR was more than doubled when applying either of the noise-reduction techniques, while a 10% increase in coverage was observed. It is worth pointing out that this percentage is an average over the 5 rats, as the coverage is dependent on the acquisition procedure such as the shim quality, the animal or the coil position.

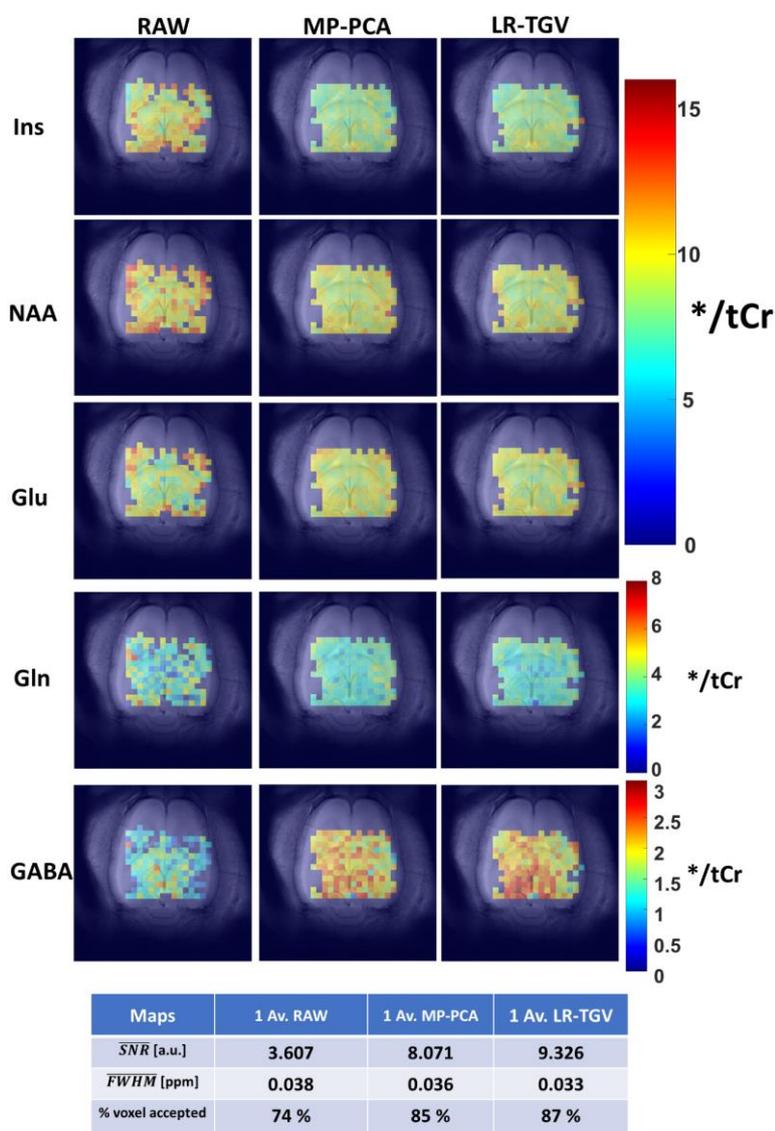

*Figure 6 In vivo concentration maps (Ins, NAA, Glu, Gln, GABA) superimposed with the anatomical slice for the three methods. The scales correspond to LCModel outputs when referenced to tCr by setting its concentration to 8 mmol/kg$_{ww}$. The table shows the average apparent SNR, FWHM and the percentage of voxel accepted for each method.*

Figure 7B shows that, regardless of the method applied, the average concentration for Gln, Glu and Ins did not change significantly while a lower SD was observed when either noise-reduction technique was applied. For the MP-PCA set, a reduction of SD from 12%, 7%, 7% to 7%, 4%, 5% was observed in the hippocampus for Gln, Glu and Ins, respectively, while a reduction of SD from 11%, 7%, 9% to 6%, 4%, 6% was observed in the striatum. For the LR-TGV set, reductions to 7%, 4%, 4% for the hippocampus and to 6%, 4%, 5% for the striatum were noted. Moreover, the brain regional difference was conserved after the application of denoising, in agreement with previous reports using non-denoised single-voxel spectroscopy (SVS) on each region[28]. Interestingly, Gln concentration has been previously reported to be higher in cortex and striatum than in hippocampus without reaching significance due to increased SD, a brain regional difference also observed in the current study with LR-TGV data showing a statistically significant difference potentially due to lower SDs following denoising. The other metabolite concentrations, relative to tCr, obtained after quantification for RAW, MP-PCA and LR-TGV can be found in Table 3.

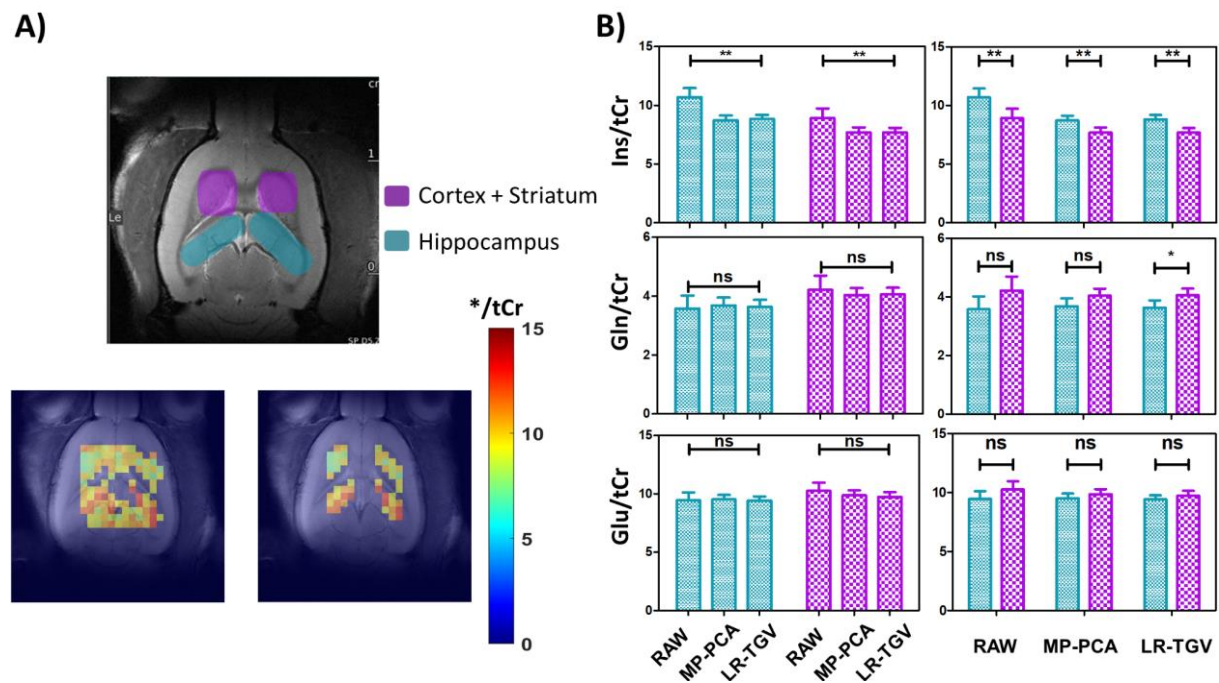

*Figure 7 A) Segmentation of the MRSI slice into two brain regions: cortex+striatum (purple) / hippocampus (blue). The scales correspond to LCModel outputs when referenced to tCr by setting its concentration to 8 mmol/kg$_{ww}$. B) Average over the number of rats (n=5) and over the selected voxels of the concentration of Ins, Gln and Glu computed for each brain region and methods used. (\*: 0.05 ≥ p / ns: not statistically significant)*

| | RAW | | | MP-PCA | | | LR-TGV | |
|---|---|---|---|---|---|---|---|---|
| */tCr [a.u] | Hippocampus | Striatum + Cortex | */tCr [a.u] | Hippocampus | Striatum + Cortex | */tCr [a.u] | Hippocampus | Striatum + Cortex |
| Ala (†) | 2.56 ± 0.25 | 3.00 ± 0.97 | Ala (†) | 1.43 ± 0.40 | 2.10 ± 0.68 | Ala (†) | 1.55 ± 0.21 | 2.08 ± 0.74 |
| Asp (†) | 2.30 ± 0.16 | 3.23 ± 1.07 | Asp (†) | 2.10 ± 0.16 | 2.45 ± 0.48 | Asp (†) | 2.24 ± 0.19 | 2.81 ± 0.26 |
| GABA (†) | 1.40 ± 0.13 | 1.43 ± 0.17 | GABA (†) | 1.94 ± 0.12 | 1.87 ± 0.14 | GABA (†) | 1.98 ± 0.08 | 1.86 ± 0.13 |
| Gln | 3.58 ± 0.44 | 4.22 ± 0.48 | Gln | 3.68 ± 0.27 | 4.04 ± 0.24 | Gln | 3.63 ± 0.25 | 4.06 ± 0.23 |
| Glu | 9.46 ± 0.66 | 10.25 ± 0.71 | Glu | 9.52 ± 0.38 | 9.87 ± 0.42 | Glu | 9.43 ± 0.34 | 9.74 ± 0.40 |
| GSH (†) | 1.65 ± 0.23 | 1.60 ± 0.10 | GSH (†) | 1.38 ± 0.13 | 1.34 ± 0.11 | GSH (†) | 1.36 ± 0.14 | 1.42 ± 0.14 |
| Ins | 10.70 ± 0.77 | 8.91 ± 0.82 | Ins | 8.72 ± 0.41 | 7.68 ± 0.44 | Ins | 8.83 ± 0.35 | 7.68 ± 0.40 |
| NAA | 9.57 ± 0.66 | 10.33 ± 0.77 | NAA | 8.87 ± 0.43 | 9.04 ± 0.44 | NAA | 8.63 ± 0.37 | 8.98 ± 0.45 |
| Tau | 8.38 ± 0.61 | 8.65 ± 0.70 | Tau | 6.83 ± 0.31 | 7.18 ± 0.34 | Tau | 7.03 ± 0.27 | 7.03 ± 0.30 |
| NAAG (†) | 2.04 ± 0.31 | 1.75 ± 0.24 | NAAG (†) | 1.16 ± 0.24 | 0.87 ± 0.17 | NAAG (†) | 0.97 ± 0.17 | 1.08 ± 0.24 |
| GPC+PCho | 1.86 ± 0.15 | 1.95 ± 0.20 | GPC+PCho | 1.65 ± 0.14 | 1.68 ± 0.14 | GPC+PCho | 1.64 ± 0.09 | 1.60 ± 0.11 |
| NAA+NAAG | 10.22 ± 0.79 | 10.82 ± 0.82 | NAA+NAAG | 9.63 ± 0.48 | 9.66 ± 0.48 | NAA+NAAG | 9.36 ± 0.43 | 9.57 ± 0.49 |

*Table 3 Average relative concentration obtained in each brain region for RAW, MP-PCA and LR-TGV (n=5). The dagger symbol (†) shows metabolites that have a relative concentration below 2 (considered lower concentrated). No significant differences were statistically measured between methods (RAW, MP-PCA and LR-TGV)*

## 4. Discussion

The goal of this work was to introduce and implement the MP-PCA and the LR-TGV denoising procedures in our processing pipeline for preclinical $^1$H-FID-MRSI. The performance of these two noise-reduction strategies was evaluated using synthetic and *in vivo* datasets, something not yet evaluated in such detail on preclinical datasets. The data structure (noise distribution and residuals map), apparent SNR and metabolite quantification (map coverage and brain regional difference precision) were used as metrics to gauge the performance of these two noise-reduction methods. With the MC simulations, similar metrics were evaluated but with different input SNR.

Both noise-reduction approaches led to similar results for MC and *in vivo* MRSI datasets. Similar characteristics of the denoised spectra were observed: noise distribution, apparent SNR gain, conserved brain regional difference. Thus, the conclusions drawn from simulations with respect to the ground truth seem to be relevant for the *in vivo* datasets.

### Noise distribution and Residuals

Our measurements on a noise-only part of the spectra in the MRSI set revealed that the Gaussian distribution was preserved for both *in vivo* and MC simulations (Figure 3). Both noise-reduction methods similarly reduced the spectral noise SD. A very specific pattern in the noise was observed for high spectral peaks such as NAA for the LR-TGV denoising, which was entirely absent on the MP-PCA where the variance of the residuals was constant throughout the signal-containing region. This pattern appeared both for simulations and *in vivo* sets and could be partially due to the effects of the $B_0$ shift mismatch between results that amplifies the residuals close to high intensity peak and near metabolite signals. Furthermore, the periodic behavior of this pattern suggested a stronger effect on either side of the water power mask, corresponding to the regions where $B_0$ shift is at its highest. It is worth noting that Clarke et al.[13] reported the presence of non-uniform variance in the metabolite region in low-rank denoising methods used in MRSI. As it stands in our study, it is hard to distinguish whether non-uniform variance or the $B_0$ shift is the main drive for such a particular pattern, but our results point towards the conclusion that $B_0$ shift distribution might have an impact on the performance of the low-rank denoising as discussed below.

## Brain coverage, SNR and linewidth

Both noise-reduction strategies increased the average LCModel apparent SNR, by more than twice its value for *in vivo* datasets and by threefold for Monte-Carlo datasets, while the linewidth for both *in vivo* and Monte-Carlo remained relatively unchanged. This increased the number of voxels passing the quality control check and consequently the brain coverage by about 10% (Figure 6). As both denoising methods did not alter linewidths, there is an intrinsic dependence between the performance of the denoising techniques (i.e. brain coverage in this case) and the shimming quality of the acquired dataset. Indeed, the maximum coverage attainable by denoising due to increased apparent SNR will be limited by a linewidth distribution map, assuming that all voxels have an acceptable RAW SNR. The new passing voxels were mostly found on the outer edges of the RAW map, which corresponds to the region where both the RF transmission and reception deteriorated due to the inhomogeneous $B_1$ profile of the surface coil, thus leading to a deterioration of RAW SNR. These voxels need to be analyzed with care, as the usual metrics for reliable quantification such as CRLB are no longer applicable when noise-reduction methods are applied[25].

## Metabolite Concentration & Regional Differences

Monte-Carlo simulations showed that, for each compartment individually, the absolute concentration for NAA and Gln was well retrieved for input SNR 5 and 12 with a consistent reduction of the SD for both denoising techniques (Figure 4). GABA showed very interesting results in synthetic data, as spatial information hidden below the noise floor in the RAW dataset at input SNR=5 were partially retrieved by both denoising techniques. In the specific case of GABA, denoising might mitigate LCModel systematic underestimation of low-concentration metabolites at low SNR. LR-TGV seemed to have merged compartments $C_1$ and $C_2$ for GABA concentration at input SNR 15 (Figure 5), and converges back to the GT at higher input SNR. Remarkably, this behavior was also observed for low-concentration metabolites such as Asp and tCho, resulting in smaller differences between brain regions (Supplementary Materials Figure S2 and S3). Because of the low-rank approximation, some signal information might be discarded and a reduction of variability (here smoothing out the concentration difference between the compartments) might be a possible consequence.[17] This effect should be studied carefully, especially when dealing with pathological versus control conditions. A possible cause for such an abrupt deviation in behavior for the LR-TGV denoising may be caused by the spatiotemporal decomposition being different for each input SNR. As the number of selected components $N_{rank}$ for the simulations was set at the value found by the MP-PCA decomposition, it is possible that this number may not be sufficient to allow LR-TGV to capture these metabolite contributions for these specific input SNR spatiotemporal decompositions. This limitation should be further tested, e.g. by manually fixing the LR-TGV to 20 components as in the *in vivo* measurement or by implementing automatic rank selection with "soft thresholding"[13].

*In vivo* measurements showed that metabolite concentration changes were not significant regardless of the dataset, implying that both denoising techniques did not alter the LCModel quantification significantly (Figure 7B). Consequently, regional differences were preserved after denoising. Metabolites that were not quantified or very poorly quantified in the RAW dataset such as Lac, PE and Glc are not shown in Table 3 due to a very low number of voxels for spatial averaging (0-6 voxels per region). It is worth noting that these metabolites were not recovered by either noise-reduction approach, which suggests that no artificial contributions of Lac, PE and Glc were introduced by MP-PCA or LR-TGV. A reduction of the SD was observed for all metabolites in Table 3. A smaller reduction in SD was noted for lower concentration metabolites such as Asp, Ala, GABA and GSH. While non-uniform variance in the metabolite region can be a factor for the LR-TGV results, similar SDs were attained with MP-PCA, which suggests that both noise-reduction methods decrease variance on lower concentration metabolites in *in vivo* settings, but to a lower extent.

### $B_0$-field map effect on denoising

One of the important distinctions between the two strategies is the use of a $B_0$-operator for the LR-TGV that is absent on the MP-PCA. On a theoretical basis, LR-TGV is a data-driven reconstruction approach, thus MRSI data should be corrected for the $B_0$ shift to reduce an unnecessary increase in the rank of the dataset. On the other hand, the MP-PCA focuses on the decomposition of the signal variance. However, results found in DW-MRS by Mosso et al[25]. showed that regardless of its presence or not in the decomposition, $B_0$ shift is important to avoid homogenization of the data, meaning that its presence leads to an increase of components that characterize the signal. In our study, a smooth, realistic $B_0$-field map (whether applied in MP-PCA or in LR-TGV) allowed for a lower number of noise eigenvalues found in the MP-PCA decomposition and a better distinction of the regions in both methods than with a randomly distributed $B_0$-field map. Moreover, it was observed through MC datasets that voxels with high $B_0$ shift values have a lower reduction of noise SD (Supplementary Materials, Figure S3). Although being a toy-model that is not representative of a complex *in vivo* situation, it does serve to show the importance of accounting for static field inhomogeneities in the denoising procedure for both techniques. *In vivo* application of either denoising techniques should be done with sets that were acquired with $B_0$ distribution as smooth and homogenous as possible.

## Limitations & Future Steps

While sufficient information and complexity were included to achieve the purpose of this study, additional features and details could be included to generate even more realistic datasets. In terms of possible artifacts, lipid and water residuals are explored as a possible constraint that could challenge both noise-reduction methods, with the implementation of a point spread function. Furthermore, in the spirit of reproducing *in vivo* conditions, a signal intensity distribution following a preclinical surface coil sensitivity profile could be added, allowing for a better understanding of voxel denoising and potential limitations of both techniques with regards to low-concentration metabolites. An interesting experimentation could be to introduce an artificial tumor/lesion in the form of a 4$^{th}$ model FID that would be a copy of either of the three model FIDs but with a specific metabolite having a different concentration. This could confront both methods to the problem of distinguishing possible pathologies in a region containing healthy voxels. For *in vivo* $^1H$-FID-MRSI data a test-retest study on a group of rodents would be of interest to investigate potential precision improvements due to denoising in *in vivo* data.

The quality control tools implemented in this version of the *$^1H$*-FID-MRSI preclinical pipeline were effective in filtering out data that were unfit for precise concentration measurements. There are, however, limitations that come with this first implementation, such as the metrics as measured by LCModel. Indeed, the LCModel SNR is defined by the ratio of the amplitude of the NAA signal and the residuals which might contain more information than noise[27]. Recently, a new version of the pipeline using in-house calculation of the FWHM and the SNR before quantification was introduced and results were fairly consistent with what was obtained with LCModel quality control criteria, allowing for a time efficient quantification. In the present study, the lowest used SNRs were in the range of LCModel SNR 4-6. As such, our study cannot speculate on any of the noise-reduction strategies using lower SNR values.

CRLB may not be a valid metric to distinguish reliable metabolite estimation for denoised datasets[13,25]. To avoid introducing any biases on the average relative concentration, the same set of voxels used for the RAW dataset were used for both MP-PCA and LR-TGV datasets. Although this approach does not take advantage of the increased coverage reported in this study, it provides an interesting test to assess the concentration SD and the preservation of regional differences.

Due to its data-driven nature, not every type of data is suited for denoising, while each denoising method can perform differently. Noise properties and information preservation should be tested before implementation. For instance, in DW-MRS, the strong variability in the data induced by different b-values led to an inhomogeneous noise level across b-values after MP-PCA denoising and imposed the usage of a sliding window[25]. For $^1H$-FID-MRSI data, this behavior was not observed, as the input datasets are inherently more redundant and feature less variability across voxels than what is observed across b-values in DW-MRS datasets. Furthermore, noise removal based on supervised DL with U-nets has been recently implemented using simulated $^1H$ MR spectra of the human brain, showing that these denoising techniques may be useful for display purposes, but do not help quantitative evaluations[35].

## 5. Conclusion

We successfully implemented two noise-reduction strategies, MP-PCA and LR-TGV, for preclinical $^1H$-FID MRSI data at 14.1T and provided a framework on how to evaluate the performance of denoising methods using synthetic data via Monte-Carlo simulations and *in vivo* data. Both denoising techniques showed an increase in apparent SNR which translated into a better MRSI spatial coverage. The relative metabolite concentration was not significantly altered by either methods and brain regional differences were preserved in both synthetic and *in vivo* datasets. SD of the estimated concentration was lowered for all metabolites. Simulations for the LR-TGV showed that compartment concentrations merged for certain SNR levels for low-concentration metabolites. Based on our results, the two denoising approaches are both suited for $^1H$-FID MRSI data at 14.1T on healthy brains, but concentration estimations ought to be treated with care especially for low-concentration metabolites. Additional tests are required to see the behavior of these denoising techniques for the study of pathological conditions. MP-PCA is a polyvalent denoising technique applicable to different types of datasets. LR-TGV is a reconstruction technique for MRSI, with denoising capabilities. Both methods also have potential for $^{31}P$ and $^2H$ MRSI data, where data are sparser than for $^1H$ MRSI, and therefore particularly suited for low rank reconstruction[36].

## Acknowledgements:

This work was supported by the Swiss National Science Foundation award n° 201218 and 207935 and by the Center for Biomedical Imaging of the UNIL, UNIGE, HUG, CHUV, EPFL, the Leenaards and Jeantet Foundations. The authors thank the veterinary staff at CIBM MRI EPFL AIT for support during experiments.

## Data Sharing:

The MATLAB code used for the generation of simulated MRSI slices is available on the following repository: https://github.com/AlvBrayan/MC_MRSI_datasetgenerator. The processing pipeline is currently being implemented in a multi-function MATLAB toolbox *MRS4Brain-toolbox* and is available on the following repository: https://github.com/AlvBrayan/MRS4Brain-toolbox. The MP-PCA MATLAB code has been made publicly available by the authors of ref.[17], on the following repository: https://github.com/NYU-DiffusionMRI/mppca_denoise. The LR-TGV MATLAB code was adapted from the authors of ref[14,37] and is available on the following repository: https://github.com/AlvBrayan/MC_MRSI_datasetgenerator

## Declaration of Conflict of Interest:

The authors have no conflict of interest to declare.

# Supplementary Materials

## 1. Data framework

The acquired signal measured with a MRSI sequence at a time t and at a Fourier-space point **k** can be expressed as:

$$s(\boldsymbol{k}, t) = \int_\Omega \rho(\boldsymbol{r}, t) e^{-2\pi i t \Delta B_0(\boldsymbol{r})} e^{-2\pi i \boldsymbol{k}\cdot\boldsymbol{r}} \, d\boldsymbol{r}$$

(1)

where $\rho(\boldsymbol{r}, t)$ is the magnetization, $\Delta B_0(\boldsymbol{r})$ is the $B_0$ magnetic field inhomogeneity profile in the slice and $\Omega \subset \mathbb{R}^3$ describes the geometric support of the system. Experimentally, the acquisition of an MRSI dataset is done in a discretized spatial and temporal grid. The acquired k-space points are described by a set of vectors $\boldsymbol{k}_i$ with $i = 1, \ldots, N_k$ and the acquired time points are described using the sampling rate in the chemical shift domain $SR$ as $t_l = \frac{l-1}{SR}$ with $l = 1, \ldots, N_T$. Similarly, the spatial coordinates are also discretized in $\boldsymbol{r}_j$ vectors ($i = 1, \ldots, N_r$) to evaluate the terms $\Delta B_0(\boldsymbol{r})$ and $\rho(\boldsymbol{r}, t)$. In the case of 2D-MRSI, it is possible to reduce (1) to a two-dimensional problem, as long as the slice are sufficiently thin to neglect partial volume effects[1]. In such a case, it is possible to express (1) as such:

$$s_{i,l} = \sum_{j=1}^{N_r} F_{i,j} B_{j,l} \rho_{j,l} + \varepsilon_{i,l}$$

(2)

with $F_{i,j} = e^{-2\pi i \boldsymbol{k}_i \cdot \boldsymbol{r}_j}$, $B_{j,l} = e^{-2\pi i t_l \Delta B_0(\boldsymbol{r}_j)}$ and $\varepsilon_{i,l}$ the additional term representing the measurement noise. It is now possible to express (2) in a more compact format which will be used for more clarity:

$$\boldsymbol{S} = \boldsymbol{\mathcal{F}}\boldsymbol{\mathcal{B}}\boldsymbol{\rho} + \boldsymbol{\varepsilon}$$

(3)

## 2. Low-rank TGV denoising

Low-rank reconstruction uses the assumption that the dataset can be partially separated within a low number of components (or rank). There are two methods of separating the MRSI dataset using low-rank: using the spatiotemporal separability of the data or using the linear predictability. For this study, the spatiotemporal assumption is used. The decomposition of the magnetization $\boldsymbol{\rho}$ using a rank $L$ can be written as such:

$$\rho_{j,l} = \sum_{n=1}^{L} U_{j,n} V_{n,l}$$

$$\boldsymbol{\rho} = \boldsymbol{UV}$$

Where $\boldsymbol{U} \in \mathbb{C}^{N_r \times L}$ and $\boldsymbol{V} \in \mathbb{C}^{L \times N_T}$ can be interpreted as the spatial and spectral components of the signal respectively. The spatiotemporal separability is justified as each metabolite resonates at a certain frequency independently of their spatial location and the noise is randomly distributed in the spatial and frequency domain and as such cannot be represented by specific spatial distribution. Effectively, the maximum rank of $\boldsymbol{\rho}$ is given by the minimum between $N_r$ and $N_T$. However, in practice, $L$ is much smaller due to the number of observable metabolites in magnetic resonance spectroscopy, enabling the use of low-rank for denoising purposes[1]. In the case of this study, low-rank decomposition is combined with a total generalized variation (TGV) constraint applied on the spatial component to avoid stair-casing artifacts as defined per Knoll et al.: $TGV^2\{U_n\} = \arg\min_{A} \ \|\nabla U_n - A\|_1 + \frac{1}{4}\|\mathcal{E}(A)\|_1$ with $\nabla$ and $\mathcal{E}$ being first and second order derivative operators respectively[2]. The optimal $\boldsymbol{U}$ and $\boldsymbol{V}$ components are then computed using a minimization problem:

$$\arg\min_{U,V} \ \|\boldsymbol{S} - \mathcal{FB}\{\boldsymbol{UV}\}\|_2^2 + \lambda \sum_{n=1}^{L} TGV^2\{U_n\}$$

Initial estimates $\boldsymbol{U_0}$ and $\boldsymbol{V_0}$ are computed from the measured signal using an adjoint operator, combined with a singular value decomposition: $\mathcal{FB}^H\{\boldsymbol{S}\} = \boldsymbol{U_0}\boldsymbol{V_0}^H$, with $\boldsymbol{U_0}$ the left singular vectors and $\boldsymbol{V_0}$ the right singular vectors multiplied by the singular value matrix.

## 3. Marchencko-Pastur PCA denoising

Marchenko-Pastur principal component analysis (MP-PCA) based denoising requires to compute the eigenvalues of the covariance matrix of $\boldsymbol{\rho_B} = \mathcal{B}\boldsymbol{\rho}^{\text{Error! Reference source not found.}}$. By applying the adjoint operator of the Fourier transform on (3), one can assume that: $\boldsymbol{\rho_B} = \mathcal{F}^H \boldsymbol{S} + \tilde{\boldsymbol{\varepsilon}}$ with $\boldsymbol{\rho_B} \in \mathbb{C}^{N_r \times N_T}$. In this study, the acquisitions were done with $2N_r < N_T$ and $2N_r \gg 1$ and as such the MP asymptotic assumptions were always met. The real and imaginary part of the temporal signal $\boldsymbol{\rho_B}$ are concatenated such that $\boldsymbol{\rho_{conc,B}} \in \mathbb{C}^{2N_r \times N_T}$. Let us define then:

$$\boldsymbol{X} = \boldsymbol{\rho_{conc,B}} - \boldsymbol{1}_{2N_r}\widehat{\boldsymbol{\rho}}_{conc,B}$$

Where $\widehat{\boldsymbol{\rho}}_{conc,B}$ is the column wise average matrix of $\boldsymbol{\rho_{conc,B}}$ of dimension $1 \times N_T$ and $\boldsymbol{1}_{2N_r}$ is column vector of ones of dimension $2N_r \times 1$. The covariance matrix is thus given by $\frac{1}{2N_r}\boldsymbol{XX}^T$ and their eigenvalues $\lambda$ can be obtained from $X$ using a singular value decomposition $X = USV^T$ as the singular values of $X$ are the square root of $\lambda$. Following the random matrix theory, it is possible to fit the Marchencko-Pastur distribution to the smallest non-zeros eigenvalues:

$$p(\lambda|\sigma, (2n-P)/m) = \begin{cases} \frac{\sqrt{(\lambda_+ - \lambda)(\lambda - \lambda_-)}}{2\pi\lambda\sigma^2(2n-P)/m} & \text{if } \lambda_- \leq \lambda \leq \lambda_+ \\ 0 & \text{otherwise} \end{cases}$$

where $\sigma$ is the noise level estimated from the input matrix $X$, $P$ is the number of signal-related eigenvalues, $\lambda_-$ the smallest noise-related eigenvalue and $\lambda_+$ the largest. $P$ corresponds to the number of values $\lambda$ such that $\lambda \geq \lambda_+$, with $\lambda_+ = \sigma^2\left(1 + \sqrt{\frac{2n}{m}}\right)^2$. The same argument about the necessary rank to describe the metabolic information as in the low-rank TGV method is applied, allowing for a denoising of the dataset. The denoised data is then computed as such:

$$\boldsymbol{\rho}^*_{conc,B} = U\, S_P V^T + \mathbf{1}_{2N_r}\widehat{\boldsymbol{\rho}}_{conc,B}$$

where, $S_P$ are the singular values truncated at the rank P.

# 4. In vivo parameters used for the study

**Supplementary Table S1:** Minimum reporting standards in MRS (Lin et al., 2021)[3]

## Hardware

| | |
|---|---|
| **Field strength** | 14.1T |
| **Manufacturer** | Bruker |
| **model (software)** | Paravision 360 V1.1. |
| **rf coil** | $^1$H-quadrature surface head coil |
| **additional hardware** | N/A |

## Acquisition

| | |
|---|---|
| **pulse sequence** | FID-MRSI |
| **volume of interest (voi)** | Rodent: Brain |
| **nominal voi size** | 0.77 × 0.77 × 2 mm$^3$ |
| **repetition time TR & Acquisition delay AD** | TR = 813ms / AD = 1.3ms |
| **number of excitations per spectrum** | 1 average |
| **additional parameters** | 24 × 24 × 2 mm$^3$ FOV; matrix size 31×31; no acceleration factor; Cartesian k-space sampling |
| **water suppression method** | VAPOR |
| **shimming method** | Bruker MAPSHIM, first in an ellipsoid covering the full brain and further in a volume of interest centered on the MRSI slice.; <30 Hz |
| **triggering or motion correction** | N/A |

## DATA ANALYSIS

| | |
|---|---|
| **Analysis Software** | LCmodel (Version 6.3-1N) |
| **Processing step deviating from reference** | Custom Basis-Set<br><br>Control files provided with the *MRS4Brain toolbox* |
| **output measure** | Ratios to total Creatine |
| **quantification reference** | Basis-set including: alanine, aspartate, ascorbate, creatine, phosphocreatine, γ-aminobutyrate, glutamine, glutamate, glycerophosphocholine, glutathione, glucose, inositol, N-acetylaspartate, N-acetylaspartylglutamate, phosphocholine, phosphoethanolamine, lactate, taurine simulated using NMR ScopeB. Macromolecules acquired in-vivo with single inversion recovery FID-MRSI |

## Data quality

| | |
|---|---|
| **reported variables** | LCModel SNR and LCModel linewidths reported |
| **Data exclusion criteria** | LCModel SNR > 0.75*LCModel and<br><br>LCModel FWHM < 1.25*LCModel |
| **quality measures of postprocessing model fitting** | CRLB < 40% |
| **sample spectrum** | Figure 2 (in vivo) / Figure 1 (Monte-Carlo) / Supplementary Figure S3-5 (Monte-Carlo RAW, MP-PCA denoised & LR-TGV denoised) |

# 5. Monte Carlo simulations - Confidence Interval optimization

This section will explain how the number of Monte-Carlo simulations $n_{MC}$ was optimized. Let us take the example of a model FID in one of the compartments (be it 2-compartments or 3-compartments model), the concentration quantified by LCModel for a specific metabolite is defined by $X_i, i = 1, \ldots, n_{MC}$. Let us define the variable $X_{n_{MC}} = \frac{1}{N}\sum_{i=1}^{n_{MC}} X_i$ corresponding to the concentration estimate averaged over the number of Monte-Carlo simulations. Using the Central Limit Theorem, with a value $n_{MC}$ sufficiently high, $X_{n_{MC}}$ follows a gaussian distribution $N(\alpha, \frac{\sigma^2}{n_{MC}})$ with a mean value $\alpha$ (corresponding to the input value of the concentration) and with a variance $\frac{\sigma^2}{n_{MC}}$ (where $\sigma$ is the noise amplitude as defined in section 2.3). The error of a Monte-Carlo simulation is defined by $\varepsilon_{n_{MC}} = \alpha - X_{n_{MC}}$ and follows a gaussian distribution centered around 0 with similar variance to $X_{n_{MC}}$. It is thus possible to rescale the error such that $\frac{\sqrt{n_{MC}}}{\sigma} \varepsilon_{n_{MC}}$ follows a normal distribution $N(0,1)$. From this result, one can express the confidence interval CI obtained through $n_{MC}$ simulation: as the probability that the term $|\frac{\sqrt{n_{MC}}}{\sigma} \varepsilon_{n_{MC}}|$ is smaller than a quantile $q$ can be estimated:

$$p(q) = \lim_{n_{MC}\to\infty} P(|\frac{\sqrt{n_{MC}}}{\sigma} \varepsilon_{n_{MC}}| < q) = \frac{1}{\sqrt{2\pi}} \int_{-q}^{q} e^{-\frac{x^2}{2}} dx$$

This equation can be interpreted as the probability $p(q)$ that our metabolite estimate averaged over the Monte-Carlo iterations $X_{n_{MC}}$ can be found in the interval $[\alpha - \frac{\sigma}{\sqrt{n_{MC}}}q ; \alpha + \frac{\sigma}{\sqrt{n_{MC}}}q]$ (Confidence Interval). The quantile $q$ is a fixed tabulated value that can be computed from a normal distribution depending on the desired two-tailed probability. In the case of this study, it was decided to use $q = 1.96$ and thus $p(q) = 95\%$. With these assumptions, the error is bounded by $\frac{\sigma}{\sqrt{n_{MC}}}q$. To compute the number of Monte-Carlo iterations needed to achieve this precision, the limit case where $\varepsilon_{n_{MC}} = \frac{\sigma}{\sqrt{n_{MC}}}q$ will be used. We define $\Delta = \alpha(1 - 0.95)$ the metabolite specific maximum deviation allowed for the Monte-Carlo simulation. The number of necessary iterations to achieve this precision with a probability of 95% is thus given by:

$$n_{MC} = \frac{1.96^2 \sigma^2}{\Delta^2}$$

For our MRSI study, we looked at regional differences between compartments. This implies a further degree of averaging corresponding to the averaging over the number of model FIDs found inside a compartment. Hence, the number of Monte-Carlo iterations needs to be divided by the number of elements found in one compartment (practically, the smallest one is chosen to set the most precise limit: $n_{comp} = 67$):

$$n_{MC} = \frac{1.96^2 \sigma^2}{n_{comp} \Delta^2}$$

From a practical standpoint, one can see that for each metabolite used for the simulation, there will be a specific number of iterations necessary. Moreover, by nature of how the Monte-Carlo simulation is built, $n_{MC}$ is dependent on the input SNR chosen. To optimize the simulation, the following assumption was applied: $n_{MC}$ is chosen with respect to the NAA for an input SNR where the LCModel SNR is sufficiently good compared to what was observed *in vivo* (in our case, input SNR = 10). These assumptions were set as such because the NAA is the most well defined peak that we observed through our RAW *in vivo* acquisition, hence its convergence in a simulated set with good SNR should be assured. Moreover, optimizing over only one metabolite allows for a more realistic approach on the lower concentrated metabolites that are more subjected to noise variation.

The first proof of concept with 2-compartment model (Supplementary Material Section 6) was produced with $n_{MC}$ = 10 (leading to $p(q) = 92\%$ for NAA at input SNR=10). For the 3-compartment model, the assumptions in the above paragraph are applied and $n_{MC}$ was found to be equal to 27, which was rounded to 30 in the present study.

# 6. Monte Carlo simulations using 2-compartments

**Supplementary Figure S1:** Concentration maps of NAA, GABA and Gln (a.u. based on concentrations reported in Supplementary Table S1), obtained from the simulated MRSI slice with the 2-compartments geometry (top of the figure), for SNR = 7 and 15. A Gaussian distributed $B_0$ field map was applied throughout the whole slice. As can be noted from the figure the concentrations in compartments C1 and C2 were swapped when comparing to Supplementary Figure S2.

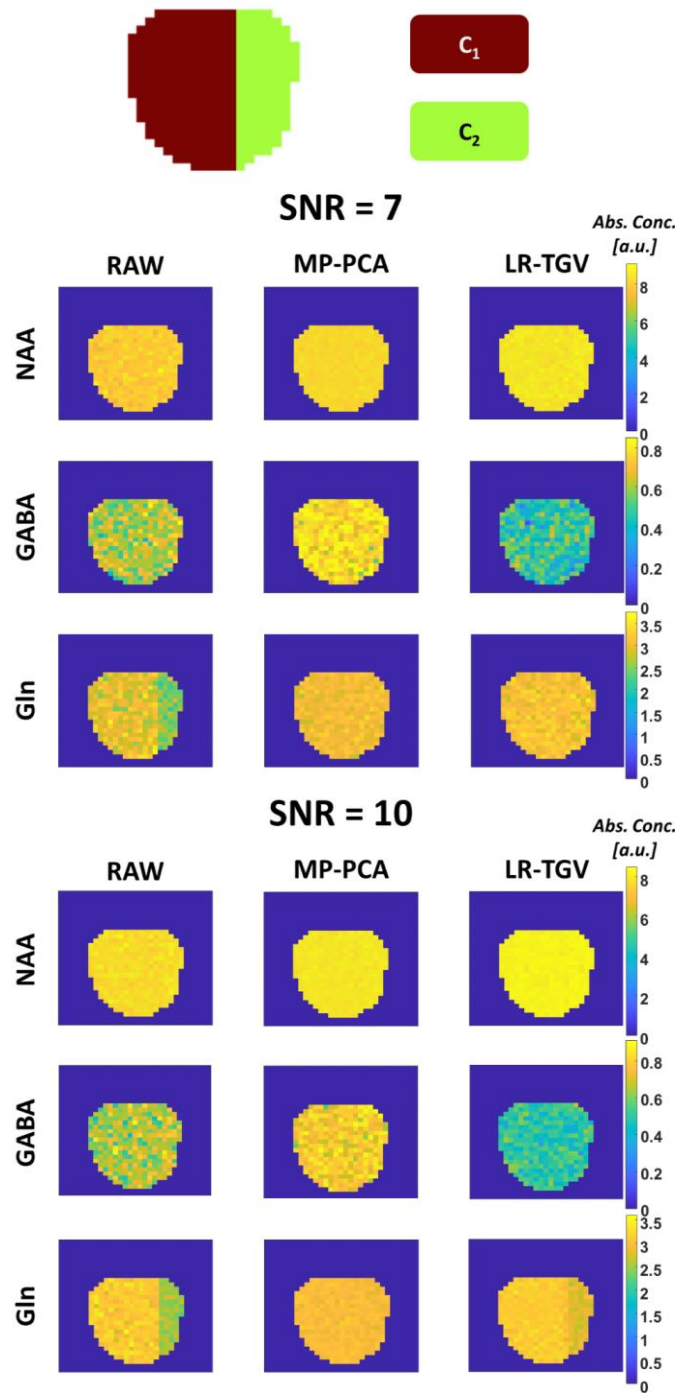

**Supplementary Figure S2:** Concentration maps of NAA, GABA and Gln (a.u. based on concentrations reported in Supplementary Table S1), obtained from the simulated MRSI slice with the 2-compartments geometry (top of the figure), for SNR = 7 and 10. The *in vivo* $B_0$ field map from Figure 1B was applied throughout the whole slice. As can be noted from the figure the concentrations in compartments C1 and C2 were swapped when comparing to Supplementary Figure S1.

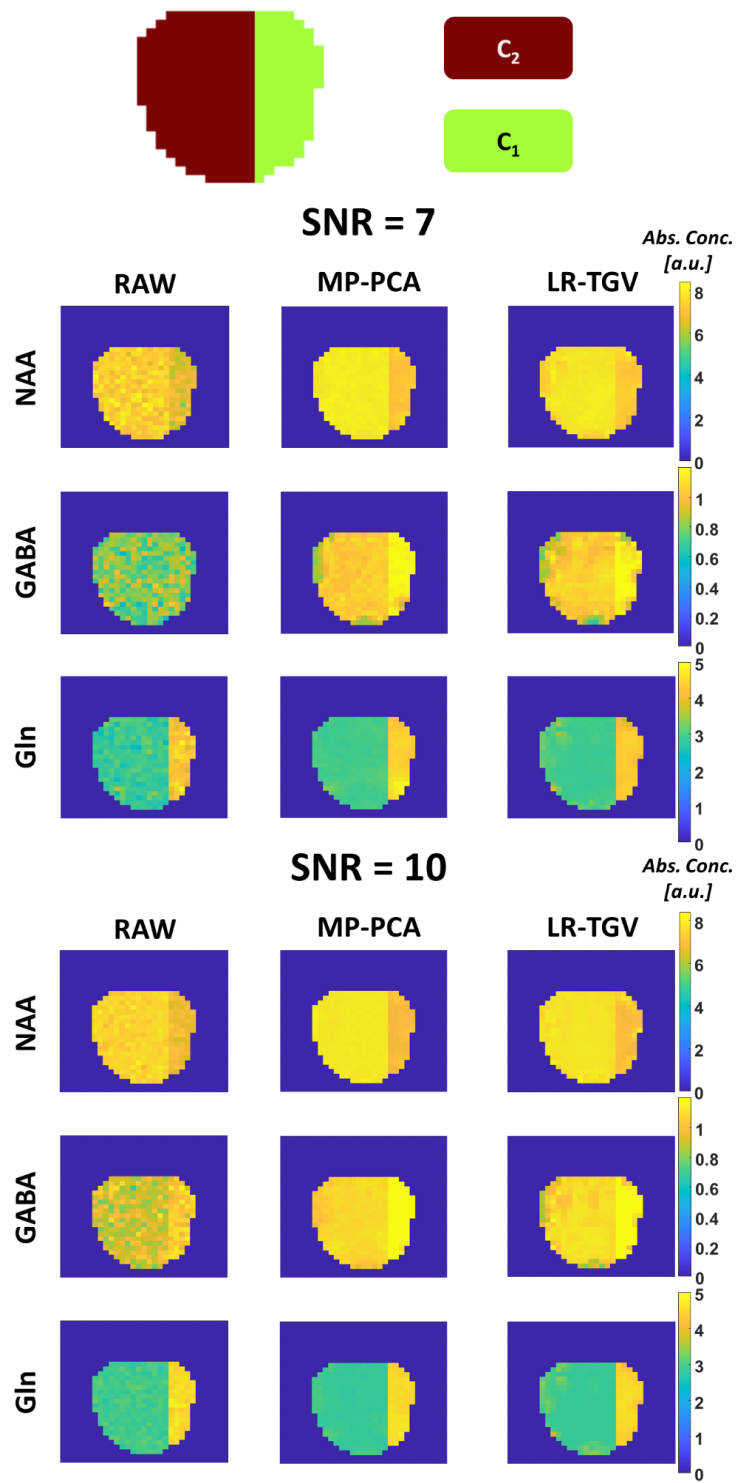

# 7. Monte-Carlo simulations using 3-compartments

**Supplementary Table S1:** Absolute concentration values used for the generation of the three model FIDs. These values were collected from *in vivo* measurements in the rodent brain[4].

| Metabolites | Compartment Concentration [a.u.] | | |
|---|---|---|---|
| | C1 | C2 | C3 |
| Ala | 0.64 | 0.67 | 0.84 |
| Asc | 2.49 | 2.04 | 2.69 |
| Asp | 1.74 | 1.6 | 1.2 |
| Cr | 4.26 | 4.03 | 5.95 |
| PCr | 4.99 | 4.46 | 6.67 |
| GABA | 1.28 | 1.44 | 1.14 |
| Gln | 3.22 | 4.79 | 3.94 |
| Glu | 10.05 | 9.68 | 10.44 |
| GSH | 0.89 | 1.06 | 0.62 |
| Ins | 7.8 | 5.19 | 6.95 |
| Lac | 1.33 | 1.44 | 1.3 |
| NAA | 9.52 | 8.1 | 9.28 |
| Tau | 6.93 | 9.74 | 5.5 |
| Glc-A | 1 | 1.1 | 1.26 |
| Glc-B | 1 | 1.1 | 1.26 |
| NAAG | 0.8 | 0.74 | 1.03 |
| PE | 2.49 | 3.42 | 1.41 |
| GPC | 0.5 | 0.53 | 0.45 |
| PCho | 0.41 | 0.83 | 0.53 |
| MM | 0.7 | 1.9 | 1.82 |

**Supplementary Figure S3:** Raw representative simulated spectra for the 3-compartments model at all input SNR values (input SNR∈{3,5,7,10,12,15,20}).

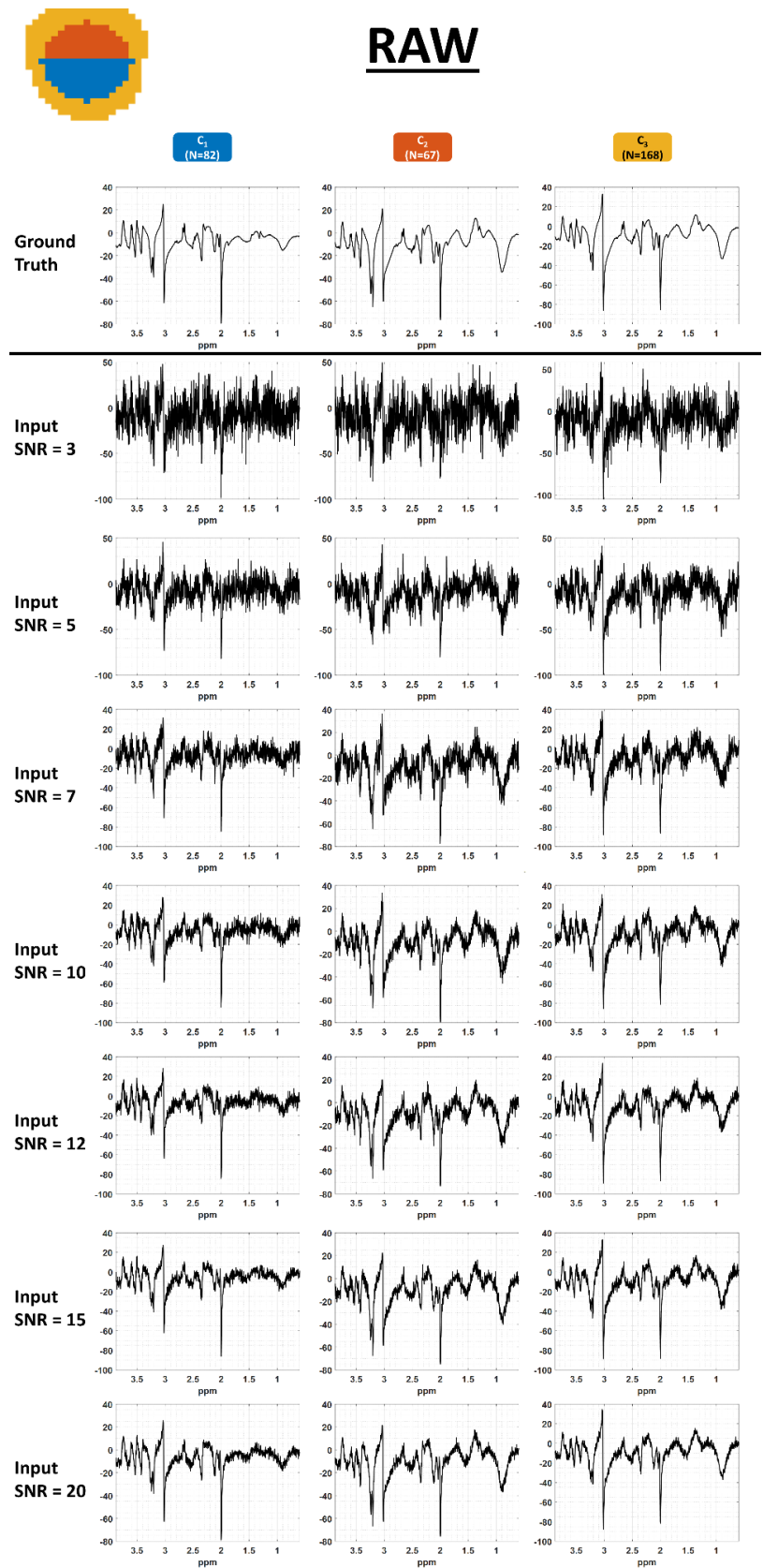

**Supplementary Figure S4:** MP-PCA denoised representative simulated spectra for the 3-compartments model at all input SNR values (input SNR∈{3,5,7,10,12,15,20}).

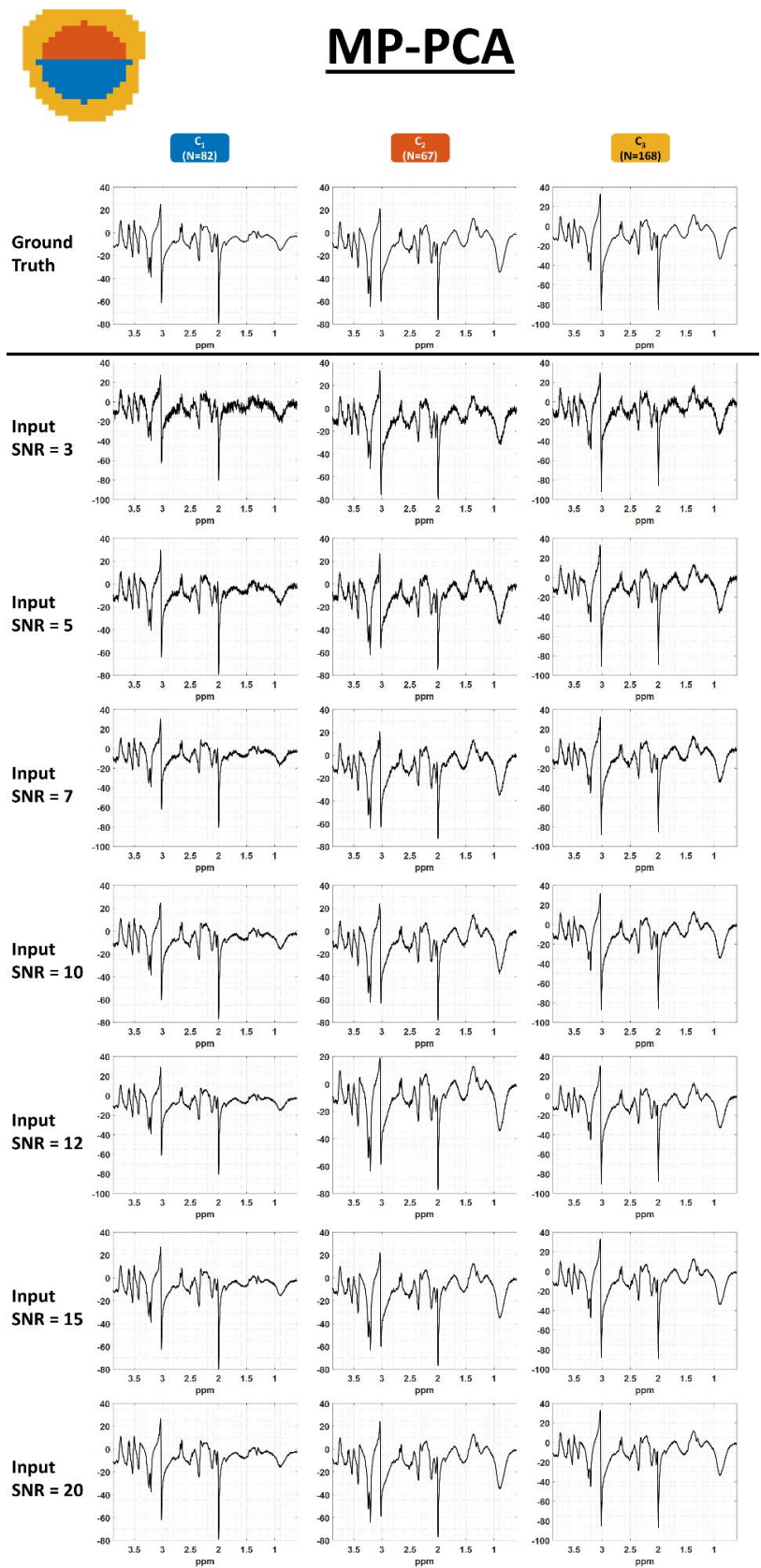

**Supplementary Figure S5:** LR-TGV denoised representative simulated spectra for the 3-compartments model at all input SNR values (input SNR∈{3,5,7,10,12,15,20}).

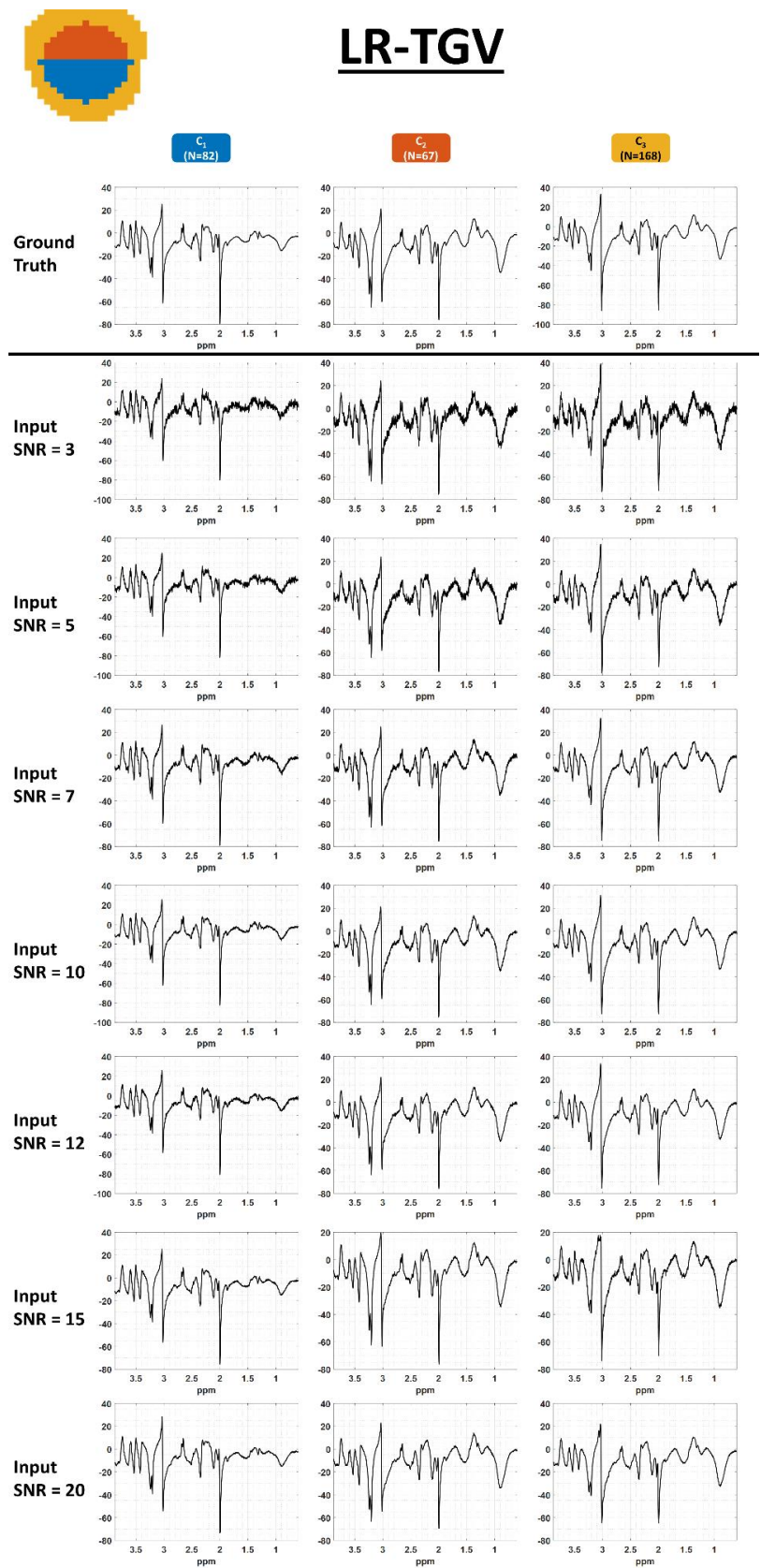

**Supplementary Figure S6:** Relative Standard Deviation noise-reduction map (%), with the corresponding $B_0$ field map (Hz), for both *in vivo* and MC dataset (input SNR=5). The noise standard deviation was computed in each voxel using a noise-only part of the real spectrum (region from 0.54 to (-0.98) ppm). The overall average noise-reduction on the map is shown in red. As can be seen the voxels in the edges of the MC maps had a smaller noise-reduction.

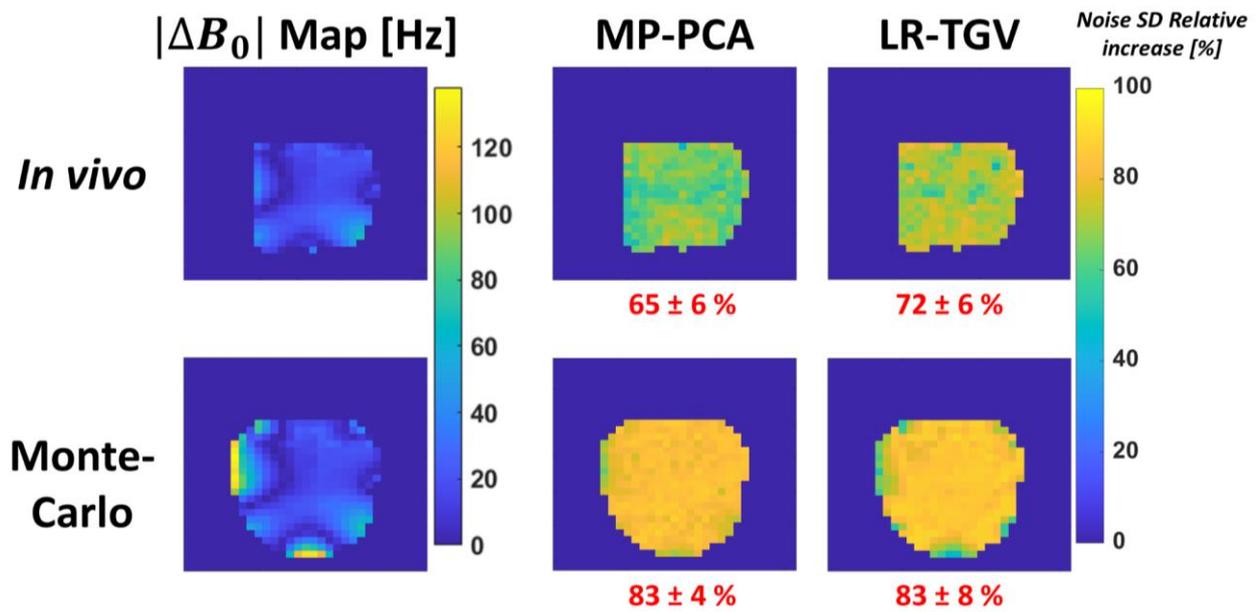

**Supplementary Figure S7:** Plots of the absolute concentration (a.u. based on concentrations reported in Supplementary Table S1) as a function of the input SNR used for the generation of the synthetic dataset. The ground truth GT value of concentration and difference is displayed in a dotted purple vertical line.

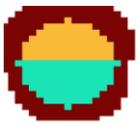
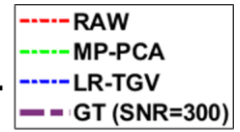
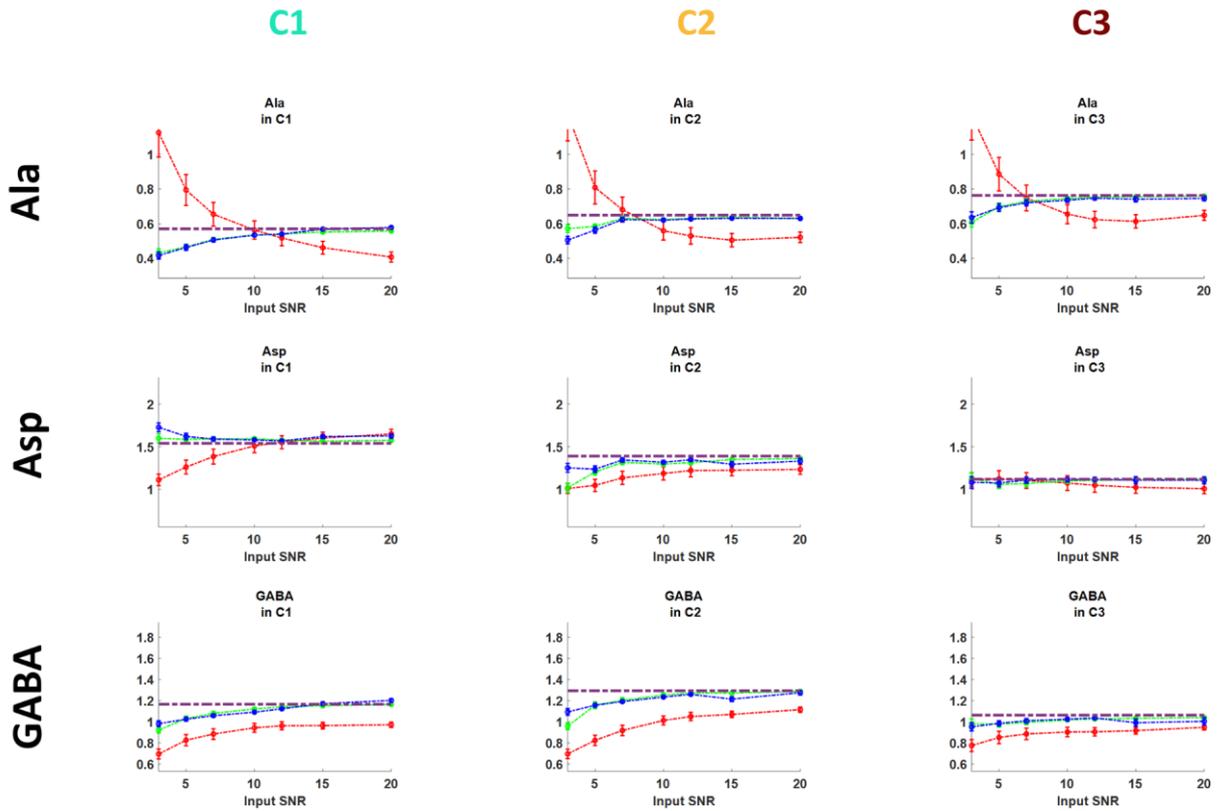

| | Gln in C1 | Gln in C2 | Gln in C3 |
| --- | --- | --- | --- |
| **Gln** | | | |

| | Glu in C1 | Glu in C2 | Glu in C3 |
| --- | --- | --- | --- |
| **Glu** | | | |

| | GSH in C1 | GSH in C2 | GSH in C3 |
| --- | --- | --- | --- |
| **GSH** | | | |

| | Ins in C1 | Ins in C2 | Ins in C3 |
| --- | --- | --- | --- |
| **Ins** | | | |

| | NAA in C1 | NAA in C2 | NAA in C3 |
| --- | --- | --- | --- |
| **NAA** | | | |

| | Tau in C1 | Tau in C2 | Tau in C3 |
| --- | --- | --- | --- |
| **Tau** | | | |

| | NAA+NAAG in C1 | NAA+NAAG in C2 | NAA+NAAG in C3 |
| --- | --- | --- | --- |
| **tNAA** | | | |

| | GPC+PCho in C1 | GPC+PCho in C2 | GPC+PCho in C3 |
| --- | --- | --- | --- |
| **tCho** | | | |

**Supplementary Figure S8:** Plots of the regional differences (in %) as a function of the input SNR (i.e. SNR of RAW data) used for the generation of the synthetic dataset. The ground truth GT value of concentration and difference is displayed in a dotted purple vertical line. The blue region represents the values of difference of 10% around the ground truth.

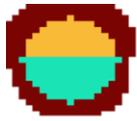
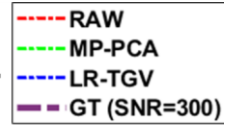
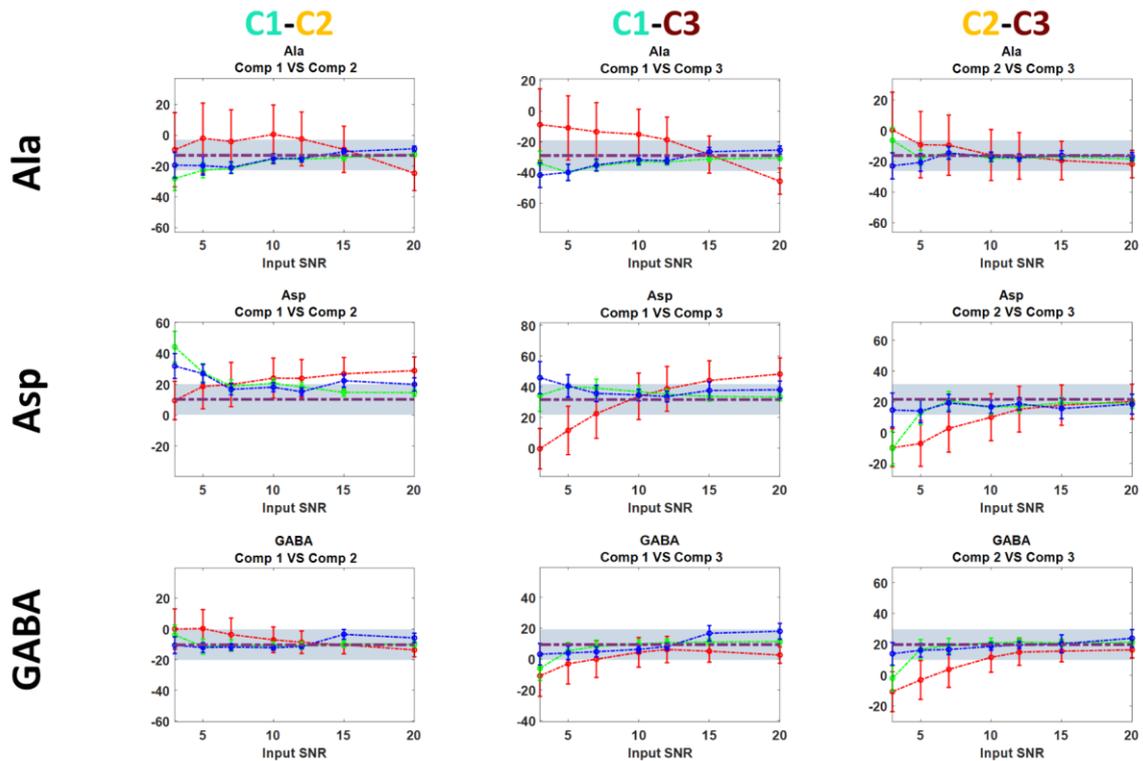

| | Comp 1 VS Comp 2 | Comp 1 VS Comp 3 | Comp 2 VS Comp 3 |
|---|---|---|---|
| **Gln** | 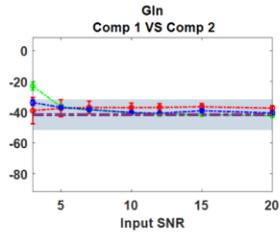 | 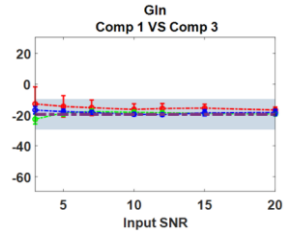 | 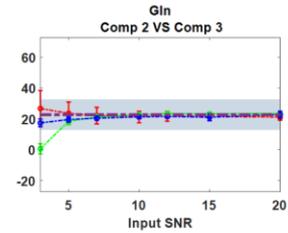 |
| **Glu** | 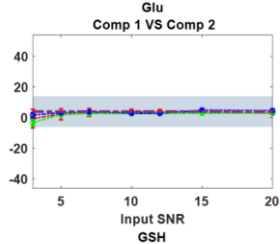 | 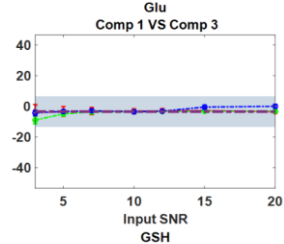 | 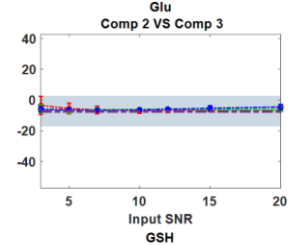 |
| **GSH** | 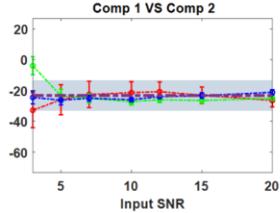 | 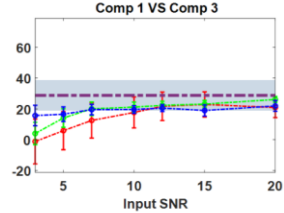 | 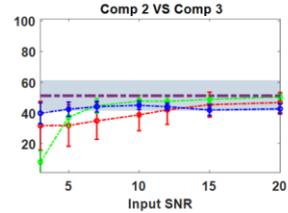 |
| **Ins** | 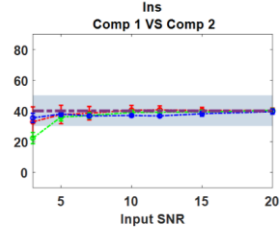 | 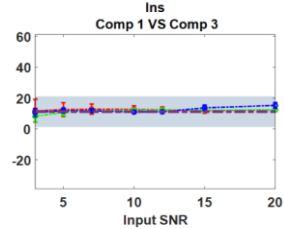 | 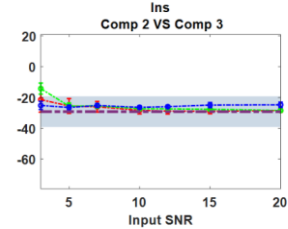 |
| **NAA** | 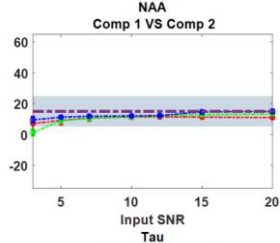 | 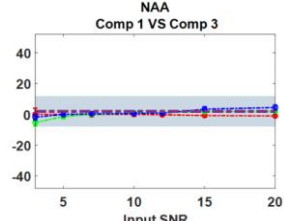 | 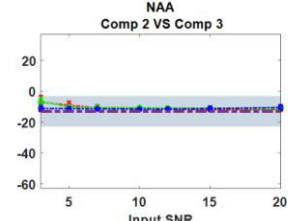 |
| **Tau** | 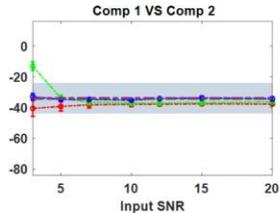 | 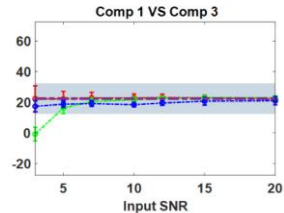 | 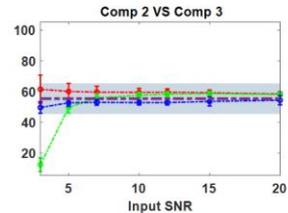 |
| **tNAA** | 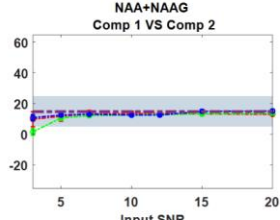 | 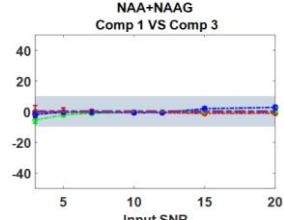 | 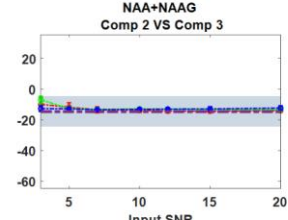 |
| **tCho** | 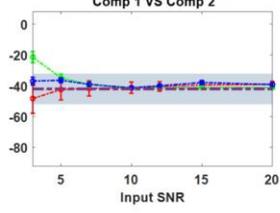 | 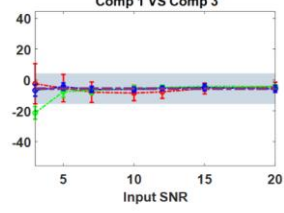 | 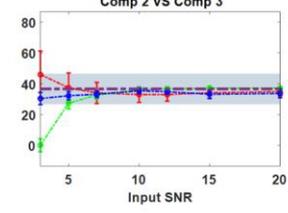 |

# References (Supplementary Materials)